\documentclass[prx,twocolumn,superscriptaddress,citeautoscript,amsart,longbibliography]{revtex4-1}

\usepackage{graphicx}
\usepackage{appendix}
\usepackage{multirow}
\usepackage{color}
\usepackage{bm}
\usepackage{times}
\usepackage{amsmath,bm,amsfonts}
\usepackage{dcolumn}
\usepackage{latexsym}

\usepackage{ulem} 
\usepackage{braket}
\usepackage{bbold}

\begin{document}
\title{Linking structures of doubly charged nodal surfaces in centrosymmetric systems}

\author{Sunje \surname{Kim}}
\affiliation{Center for Correlated Electron Systems, Institute for Basic Science (IBS), Seoul 08826, Korea}
\affiliation{Department of Physics and Astronomy, Seoul National University, Seoul 08826, Korea}
\affiliation{Center for Theoretical Physics (CTP), Seoul National University, Seoul 08826, Korea}

\author{Dong-Choon \surname{Ryu}}
\affiliation{Center for Correlated Electron Systems, Institute for Basic Science (IBS), Seoul 08826, Korea}
\affiliation{Department of Physics and Astronomy, Seoul National University, Seoul 08826, Korea}
\affiliation{Center for Theoretical Physics (CTP), Seoul National University, Seoul 08826, Korea}

\author{Bohm-Jung \surname{Yang}}
\email{bjyang@snu.ac.kr}
\affiliation{Center for Correlated Electron Systems, Institute for Basic Science (IBS), Seoul 08826, Korea}
\affiliation{Department of Physics and Astronomy, Seoul National University, Seoul 08826, Korea}
\affiliation{Center for Theoretical Physics (CTP), Seoul National University, Seoul 08826, Korea}

\date{\today}

\begin{abstract}
In topological semimetals and nodal superconductors, band crossings between occupied and unoccupied bands form stable nodal points/lines/surfaces carrying quantized topological charges. In particular, in centrosymmetric systems, some nodal structures at the Fermi energy $E_F$ carry two distinct topological charges, and thus they are called doubly charged nodes. 
Here we show that doubly charged nodal surfaces of centrosymmetric systems in three-dimensions always develop peculiar linking structures with nodal points or lines formed between occupied bands below $E_F$. 
Such linking structures can naturally explain the inherent relationship between the charge of the node below $E_F$ and the two charges of the nodal surfaces at $E_F$.
Based on the Altland-Zirnbauer (AZ)-type ten-fold classification of nodes with additional inversion $\mathcal{I}$ symmetry, which is called the AZ$+\mathcal{I}$ classification, we provide the complete list of linking structures of doubly charged nodes in centrosymmetric systems. The linking structures of doubly charged nodes clearly demonstrate that not only the local band structure around the node but also the global band structure play a critical role in characterizing gapless topological phases.
\end{abstract}

\maketitle

{\it Introduction.---}
Gapless topological states, such as topological semimetals or nodal superconductors, host stable nodal points/lines/surfaces
at the Fermi energy $E_F$. 
The stability of a node is normally characterized by a primary topological charge, defined in a lowest dimensional manifold enclosing the node in momentum space. 
For instance, in three-dimensional (3D) systems, the primary topological charges of nodal points (NPs), nodal lines (NLs), nodal surfaces (NSs) 
are defined in a two-dimensional (2D) \cite{Morimoto2014WSMDSMcharge,vafek2014dirac,xu2011chern,
delplace2012topological,fang2012multi,wan2011topological,
burkov2011weyl,armitage2018weyl,sau2012topologically,
balents2012weyl,Zhao2017PTSMin2D,balents2011TNS,
Zhao2016TSTCS,bzdusek2018convWPandNL,sigrist2018SCwithoutTandI,
wang2019higher,yanase2019classifTCSCnode}, one-dimensional (1D) \cite{bzduvsek2017doubly,weng2015topological,sato2006nodal,
Fang2015PTNLSM,Fang2019diagNLwithRot,beri2010topologically,
Song2018nonmagneticTS,Zhao2017PTSMin2D,Takahashi2017hourglass,
balents2011TNS,bzduvsek2019nonabelian,bzduvsek2019linking,
ahn2018linking,bzdusek2018convWPandNL,sim2019dsSC,wang2019higher,
yanase2019classifTCSCnode,ahn2019SWclass,Shingo2018linenodeSC,
Shingo2016linenodeSC,Shingo2014linenodeSC}, zero-dimensional (0D) \cite{wu2018nodalsurface,bzduvsek2017doubly,
oh2020instabofBogFS,turker2018NS,brydon2018bogFS,
xiao2020chargedNS,sim2019dsSC,lapp2020expeofBogFS,
volkov2018coulindinstaNS} subspaces enclosing the node, respectively. 
However, except the case of NPs, the primary topological charges of NLs and NSs do not guarantee their global stability.
Namely, a NL (NS) with a nontrivial 1D (0D) topological charge can be continuously deformed to a point and then be annihilated.
In this respect, the 1D (0D) primary topological charge of a NL (NS) indicates only local stability of a part of the node embraced by the enclosing subspace.

\begin{figure}[t]
\includegraphics[width=\linewidth]{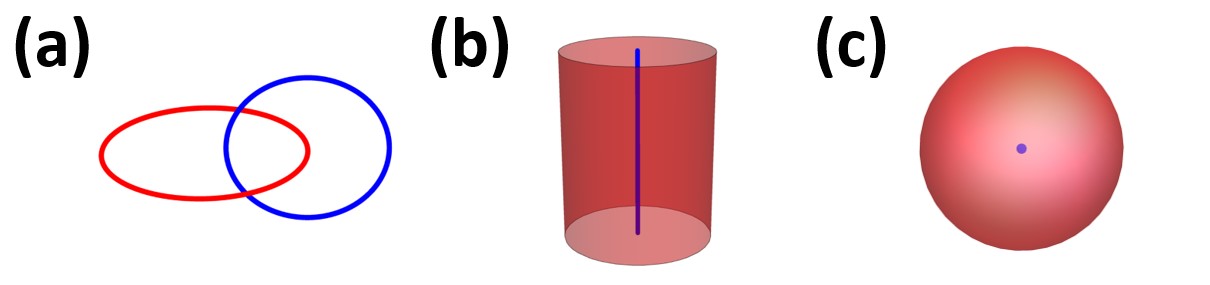}
\caption{
All possible linking structures of nodes in 3D centrosymmetric systems. Red (Blue) color indicates the node at the Fermi energy $E_F$ (below $E_F$).
(a) A nodal line (NL) at $E_F$ is linked with another NL below $E_F$ in class AI and CI. 
(b) A nodal surface (NS) at $E_F$ is linked with a NL below $E_F$ in class BDI. 
(c) A NS at $E_F$ is linked with a nodal point (NP) below $E_F$ in class D.}
\label{fig:geolink}
\end{figure}

However, according to the recent studies of 3D NL semimetals in spinless fermion systems with inversion $\mathcal{I}$ and time-reversal $\mathcal{T}$ symmetries, there is a class of NLs which are much more robust than ordinary NLs~\cite{Fang2015PTNLSM, ahn2018linking}.
Such a robust NL carries not only a 1D primary topological charge but also a 2D monopole charge, and thus it is {\it doubly charged}~\cite{bzduvsek2017doubly}.
A doubly charged NL (DCNL) with a nonzero monopole charge, called a monopole NL, is stable and cannot be annihilated as long as it does not merge with another monopole NL.
Moreover, a monopole NL at $E_F$ is always linked with other NLs below $E_F$ developing so-called the {\it linking structure} [see Fig.~\ref{fig:geolink} (a)].
The extra stability of DCNLs can be naturally explained when such a linking structure is considered~\cite{bzduvsek2019linking,ahn2018linking}.

Doubly charged nodes can also appear in the form of NSs.
Recently, for instance, by extending the Altland-Zirnbauer (AZ) classification of topological states to the cases with additional inversion $\mathcal{I}$ symmetry, 
a systematic classification of nodal structures, called the AZ$+\mathcal{I}$ classification, was performed~\cite{bzduvsek2017doubly}. It is found that doubly charged nodes can exist in four different AZ$+\mathcal{I}$ symmetry classes. Namely, the class AI and CI support DCNLs while the class BDI and D support doubly charged NSs (DCNSs). Also a recent study has shown that a DCNL in the class CI develops a linking structure~\cite{bzduvsek2019linking}, which is similar to monopole NLs belonging to class AI~\cite{ahn2018linking}. 
Considering the close relationship between the doubly charged nature of NLs and the corresponding linking structures, it is natural to expect that similar linking structures 
can also be developed in systems with DCNSs.

In this Letter, we show that DCNSs at $E_F$ of 3D centrosymmetric systems, belonging to class BDI, exhibit unusual linking structures with NLs below $E_F$ 
[see Fig.~\ref{fig:geolink}(b)]. 
The presence of NPs inside DCNSs in class D discovered in~\cite{bzduvsek2017doubly}, can also be understood in terms of linked nodal structures [see Fig.~\ref{fig:geolink}(c)].   
In these systems, the linking structure naturally explains the fundamental relationship between the primary charge of nodes below $E_F$ and the two charges of the NS at $E_F$. 
Based on the AZ$+\mathcal{I}$ classification, we provide the complete list of linking structures of doubly charged nodes in centrosymmetric systems in Table~\ref{table:charge}.

{\it AZ$+\mathcal{I}$ classification of nodes.---}
Let us first briefly recap the idea of the AZ$+\mathcal{I}$ classification. 
The standard AZ classification~\cite{altland1997nonstandard} classifies gapped band structures with time reversal $\mathcal{T}$, particle-hole $\mathcal{P}$, and chiral $\mathcal{C}$ symmetries.  
On the other hand, the AZ+$\mathcal{I}$ classification~\cite{bzduvsek2017doubly} investigates stable nodes located at generic momentum $\mathbf{k}$ in $\mathcal{I}$ symmetric systems, based on the following three $\mathbf{k}$-local symmetries: $\mathfrak{T}\equiv\mathcal{TI}$, $\mathfrak{B}\equiv\mathcal{PI}$, and $\mathcal{C}$, which transform the Hamiltonian $H(\mathbf{k})$ as
$\mathfrak{T}H(\mathbf{k})\mathfrak{T}^{-1}=H(\mathbf{k})\label{PT}$,
$\mathfrak{B}H(\mathbf{k})\mathfrak{P}^{-1}=-H(\mathbf{k})\label{PB}$,
$\mathcal{C}H(\mathbf{k})\mathcal{C}^{-1}=-H(\mathbf{k})\label{C}$,
where $\mathfrak{T}^2=\mathcal{T}^2$ since $\mathcal{T}$ and $\mathcal{I}$ commute, while $\mathfrak{B}^2=\pm\mathcal{P}^2$ depending on the commutation relation between $\mathcal{P}$ and $\mathcal{I}$. For superconductors, $H(\mathbf{k})$ indicates the Bogoliubov-de Genn (BdG) mean-field Hamiltonian.

\begin{table}[t]
\begin{tabular}{p{0.09\textwidth}<{\centering}|p{0.05\textwidth}<{\centering}p{0.05\textwidth}<{\centering}p{0.05\textwidth}<{\centering}|p{0.1\textwidth}<{\centering}}
\hline
\hline
\multirow{2}{*}{AZ$+\mathcal{I}$ class}& \multicolumn{3}{c|}{$\pi_{p}(M_{H})$} & \multirow{2}{*}{node type}\\
&  p=0 & p=1 & p=2 & \\
\hline
\multirow{2}{*}{AI}  & $\mathbb{0}$ & \textcolor{red}{$\mathbb{Z}_{2}$} & $\pmb{\mathbb{Z}_{2}}$ & line\\
&  $\mathbb{0}$ & \textcolor{red}{$\mathbb{Z}_{2}$} & $\mathbb{Z}_{2}$ & line\\
\hline
\multirow{2}{*}{BDI}  & \textcolor{red}{$\mathbb{Z}_{2}$} & $\pmb{\mathbb{Z}_{2}}$ & $\mathbb{0}$ & surface \\
&  $\mathbb{0}$ & \textcolor{red}{$\mathbb{Z}_{2}$} & $\mathbb{Z}_{2}$ & line \\
\hline
\multirow{2}{*}{D}  & \textcolor{red}{$\mathbb{Z}_{2}$} & $\mathbb{0}$ & $\pmb{2\mathbb{Z}}$ & surface\\
&  $\mathbb{0}$ & $\mathbb{0}$ & \textcolor{red}{$\mathbb{Z}$} & point\\
\hline
\multirow{2}{*}{CI}  & $\mathbb{0}$ & \textcolor{red}{$\mathbb{Z}$} & $\pmb{\mathbb{Z}_{2}}$ & line\\
&  $\mathbb{0}$ & \textcolor{red}{$\mathbb{Z}_{2}$} & $\mathbb{Z}_{2}$ & line\\
\hline
\hline
\end{tabular}
\caption{Topological charges of doubly charged nodes at $E_F$ and related nodes below $E_F$. 
The second column denotes the homotopy class of the classifying space $M_{H}$ when the numbers of occupied and unoccupied bands are large.
Here, for each class, the upper (lower) row indicates the topological invariant of a node at $E_F$ (below $E_F$) for a given dimension $p$ of the manifold enclosing the node. 
In each class, the bold dark symbol indicates the higher dimensional topological charge of a doubly charged node at $E_F$ while the red symbols denote the primary topological charges of the nodes at and below $E_F$ which are related to the bold dark symbol through the linking structure. 
The last column indicates the node shape where the upper (lower) one corresponds to the node at $E_F$ (below $E_F$).
}
\label{table:charge}
\end{table}

The topological charge of a node is given by the $p$-th homotopy group $\pi_{p}$ ($p=0,~1,~2$) of $H(\mathbf{k})$ (or its classifying space $M_{H}$). Depending on the properties of $\mathfrak{T}$, $\mathfrak{B}$, $\mathcal{C}$ symmetries, 
we have different homotopy invariants as shown in Table~\ref{table:charge}. 
The nodes at and below $E_F$ generally belong to different AZ$+\mathcal{I}$ classes.
Explicitly, the nodes below $E_F$ for the class AI, BDI, CI have only $\mathfrak{T}$ symmetry satisfying $\mathfrak{T}^2=+1$ and belong to the class AI while those for the class D has no symmetry and belongs to the class A. The corresponding topological charges are also shown in Table~\ref{table:charge}. 
Interestingly, for each class in Table I, the dimension of the secondary charge of the node at $E_F$ (bold symbol in Table I) is identical to the sum of the dimensions of the primary charges (red symbols in Table I) for two nodes at and below $E_F$, respectively, which is in accordance with the intrinsic linking between the relevant nodal structures shown in Fig.~\ref{fig:geolink}.
For instance, it was shown recently that the 2D topological charge of NLs in the class AI and CI is given by the product of the primary charge of the NLs  at and below $E_F$~\cite{ahn2018linking,bzduvsek2019linking} [see Fig.~\ref{fig:geolink} (a)].
Below we unveil such an intriguing relationship for DCNSs shown in Fig.~\ref{fig:geolink} (b,~c).

{\it Nodal structure in class BDI.---}
In class BDI, we have $\mathfrak{T}^2=\mathfrak{B}^2=\mathcal{C}^2=1$.
Suppose that there are $N$ occupied and $N$ unoccupied bands, and the energies of unoccupied bands are labelled as
$0\leq E_{1\mathbf{k}} \leq \cdots \leq E_{N\mathbf{k}}$.
Since $\{H(\mathbf{k}), \mathcal{C}\}=0$, for an occupied state $|u_{n\mathbf{k}}^{\text{occ}}\rangle$ with the energy $-E_{n\mathbf{k}}$,
there is a relevant unoccupied state $|u_{n\mathbf{k}}^{\text{unocc}}\rangle\propto \mathcal{C} |u_{n\mathbf{k}}^{\text{occ}}\rangle$ with the energy $E_{n\mathbf{k}}$ ($n=1,\cdots, N$). 
For convenience, we take the following symmetry representation $\mathfrak{T}=\mathcal{K}$, $\mathcal{C}=\sigma_{z}$, and $\mathfrak{B}=\mathfrak{T}\mathcal{C}$ where $\mathcal{K}$ indicates the complex conjugation operator, and the Pauli matrices $\sigma_{x,y,z}$ act on the particle-hole space. 
Due to $\mathcal{C}=\sigma_z$ symmetry, $H(\mathbf{k})$ takes a block off-diagonal form as $H(\mathbf{k})=\begin{pmatrix} 0&A(\mathbf{k})\\A^{T}(\mathbf{k})&0 \end{pmatrix}$ where $A(\mathbf{k})$ denotes a $N\times N$ real matrix. Also $|u_{n\mathbf{k}}^{\text{occ}}\rangle$ and $|u_{n\mathbf{k}}^{\text{unocc}}\rangle$ can be chosen as
\begin{gather}
|u_{n\mathbf{k}}^{\text{occ}}\rangle ={1 \over \sqrt{2}} \begin{pmatrix}
|u_{n\mathbf{k}}^{\uparrow}\rangle \\ |u_{n\mathbf{k}}^{\downarrow}\rangle
\end{pmatrix},\; 
|u_{n\mathbf{k}}^{\text{unocc}}\rangle ={1 \over \sqrt{2}} \begin{pmatrix}
|u_{n\mathbf{k}}^{\uparrow}\rangle \\ -|u_{n\mathbf{k}}^{\downarrow}\rangle
\end{pmatrix},
\label{BDIstate}
\end{gather}
where $|u_{n\mathbf{k}}^{\uparrow}\rangle\left(|u_{n\mathbf{k}}^{\downarrow}\rangle\right)$ are $N$-component vectors that satisfy
$\langle u^{\uparrow}_{n\mathbf{k}}|u^{\uparrow}_{m\mathbf{k}} \rangle=\langle u^{\downarrow}_{n\mathbf{k}}|u^{\downarrow}_{m\mathbf{k}} \rangle=\delta_{nm}$.
Then $A(\mathbf{k})$ can be written as $A(\mathbf{k})=\sum_{n=1}^{N} E_{n\mathbf{k}}|u^{\uparrow}_{n\mathbf{k}}\rangle\langle u^{\downarrow}_{n\mathbf{k}}|$.
A band inversion between $|u^{\mathrm{occ}}_{1\mathbf{k}}\rangle$ and $|u^{\mathrm{unocc}}_{1\mathbf{k}}\rangle$ generates a cylindrical NS at $E_F$,
and the corresponding $A(\mathbf{k})$ is given by
\begin{gather}
A^{\pm}(\mathbf{k})=\pm E_{1\mathbf{k}}|u^{\uparrow}_{1\mathbf{k}}\rangle\langle u^{\downarrow}_{1\mathbf{k}}|+\sum_{n=2}^{N} E_{n\mathbf{k}}|u^{\uparrow}_{n\mathbf{k}}\rangle\langle u^{\downarrow}_{n\mathbf{k}}|,
\label{eq:A_pm}
\end{gather} 
where $A^+(\mathbf{k})$ ($A^-(\mathbf{k})$) corresponds to the Hamiltonian inside (outside) of the NS.

The nodal structure at $E_F$ can generally be described by an effective two-band Hamiltonian $H_{\text{eff}}(\mathbf{k})=h_x\sigma_x+h_y\sigma_y+h_z\sigma_z$ with real functions $h_{x,y,z}(\mathbf{k})$ spanned by $|u_{1\mathbf{k}}^{\text{occ}}\rangle$ and $|u_{1\mathbf{k}}^{\text{unocc}}\rangle$. As $\mathfrak{T}$ and $\mathcal{C}$ symmetries require $h_y=h_z=0$, the energy gap can be closed if and only if $h_x(\mathbf{k})=0$. Since there are three momentum variables while only one equation needs to be satisfied,
a NS is expected at $E_F$. In the case of nodes below $E_F$, as the relevant Hamiltonian $H_{\text{eff}}(\mathbf{k})$ spanned by $|u_{1,2\mathbf{k}}^{\text{occ}}\rangle$ has only $\mathfrak{T}$ symmetry that gives $h_y=0$, NLs are expected.
Generally, in class BDI, NSs (NLs) appear at (below) $E_F$.

Explicitly, the topological charges of the NS at $E_F$ are defined as follows.
The 0D charge $c_{\mathrm{0D}}$ is defined as
\begin{align}
c_{\mathrm{0D}}=\mathrm{sign}\{\mathrm{det}A(\mathbf{k}_{\text{in}})\cdot \mathrm{det}A(\mathbf{k}_{\text{out}})\},
\label{0D_BDI}
\end{align}
where $\mathbf{k}_{\text{in}}$ ($\mathbf{k}_{\text{out}}$) indicates a momentum inside (outside) the NS~\cite{bzduvsek2017doubly}. From Eq.~(\ref{eq:A_pm}), we obtain $c_{\mathrm{0D}}=-1$ for a NS obtained by a band inversion.   

To define the 1D charge $c_{\mathrm{1D}}$ of the NS, we consider the spectral flattening $E_{1\mathbf{k}}=E_{2\mathbf{k}}=\cdots=E_{N\mathbf{k}}=1$. 
Then 
\begin{gather}
c_{\mathrm{1D}}\equiv\left[A_{FB}:S^{1}\rightarrow\mathrm{O(N)}\right],
\end{gather}
where $A_{FB}$ is an off-diagonal block of the flattened Hamiltonian which is an element of $\mathrm{O(N)}$.
$S^{1}$ is a circle encircling the NS.
$\left[A_{FB}:S^{1}\rightarrow\mathrm{O(N)}\right]$ means the homotopy equivalence class of $\mathrm{O(N)}$ group \cite{bzduvsek2017doubly}.
 
The 1D charge $\tilde{c}_{\mathrm{1D}}$ of a NL below $E_F$ is defined as follows.
 For a NL formed between $|u_{n,\mathbf{k}}^{\mathrm{occ}}\rangle$ and $|u_{n+1,\mathbf{k}}^{\mathrm{occ}}\rangle$, we consider a circle enclosing it, parametrized by $\theta\in[-\pi,\pi]$.
 Assuming that $|u_{i,\mathbf{k}}^{\mathrm{occ}}\rangle$ $(i=n, n+1)$ changes continuously for $\theta\in(-\pi,\pi)$ and 
 taking the representation $\mathfrak{T}=\mathcal{K}$ so that the eigenstates become real-valued,
 the following should hold at $\theta=\pm \pi$~\cite{ahn2018linking},
\begin{gather}
|u_{i,\pi}^{\mathrm{occ}}\rangle=\pm|u_{i,-\pi}^{\mathrm{occ}}\rangle.
\label{eq:1stSW}
\end{gather}
If $|u_{i,\mathbf{k}}^{\mathrm{occ}}\rangle$ changes discontinuously (continuously) at $\theta=\pm \pi$,  
 $\tilde{c}_{\mathrm{1D}}$ is non-trivial (trivial)~\cite{ahn2018linking}.

{\it Linking structure.--} 
Linking structure arises if $c_{\mathrm{1D}}=\tilde{c}_{\mathrm{1D}}$ for a NS with $c_{\mathrm{0D}}\neq0$,
because $\tilde{c}_{\mathrm{1D}}\neq0$ only when a NL exists below $E_F$.
To prove this, we define $A^{\pm}_{FB}(\mathbf{k})$ corresponding to $A^{\pm}(\mathbf{k})$ in Eq.~(\ref{eq:A_pm}) for flattened Hamiltonians.
Also, we additionally define $A'_{FB}(\mathbf{k})$ which has the same form as $A^{-}_{FB}(\mathbf{k})$ but defined inside the NS, 
thus it is irrelevant to the physical Hamiltonian.

To evaluate $c_{\textrm{1D}}$, let us consider a circle $S^{1}$ surrounding the NS.
$c_{\textrm{1D}}$ is given by the homotopy equivalence class of $A^{-}_{FB}(\mathbf{k})$ on $S^{1}$.
As $A^{-}_{FB}(\mathbf{k})$ defined outside the NS is continuously connected with $A'_{FB}(\mathbf{k})$ defined inside the NS,
$c_{\textrm{1D}}$ can be equivalently described by $A'_{FB}(\mathbf{k})$ as 
$c_{\textrm{1D}}=\left[A'_{FB}:S'^{1}\rightarrow \mathrm{O(N)}\right]$
where $S'^{1}$ is a circle inside the NS obtained by deforming $S^{1}$ continuously.
Then we ask whether $S'^{1}$ can be shrunk to a point while keeping $A'_{FB}$ well-defined on it. Generally, such a smooth deformation is impossible when there is a NL inside $S'^{1}$ where $|u^{\mathrm{occ}}_{1\mathbf{k}}\rangle$ and $|u^{\mathrm{occ}}_{2\mathbf{k}}\rangle$ are degenerate so that $A'_{FB}$ cannot be well defined.
When $S'^{1}$ encircling the NL is sufficiently shrunk, $A'_{FB}(\mathbf{k})$ may either stay nearly constant or oscillate prominently on $S'^{1}$.
In the former (latter) case, the homotopy equivalence class of $A'_{FB}$ defined on $S'^{1}$ is trivial (non-trivial).
This information is sufficient to characterize $c_{\textrm{1D}}$ because $\pi_{1}(\mathrm{O(N)})=\mathbb{Z}_{2}$ for $\mathrm{N}>2$.
For $\mathrm{N}=2$, as $\pi_{1}(\mathrm{O(2)})=\mathbb{Z}$, the relevant integer winding number should be explicitly computed as shown below.

$c_{\mathrm{1D}}=\tilde{c}_{\mathrm{1D}}$ can be shown as follows.
Since $A^{+}_{FB}(\mathbf{k})$ is well-defined inside the NS, $c_{\textrm{1D}}$ can also be determined from $A^{+}_{FB}(\mathbf{k})-A'_{FB}(\mathbf{k})=2|u^{\uparrow}_{1\mathbf{k}}\rangle\langle u^{\downarrow}_{1\mathbf{k}}|$ on $S'^{1}$, instead of $A'_{FB}(\mathbf{k})$.
From Eq.~(\ref{eq:1stSW}), we can show that, when $\tilde{c}_{\textrm{1D}}$ is nontrivial,
$|u^{\uparrow}_{1\mathbf{k}}\rangle\langle u^{\downarrow}_{1\mathbf{k}}|$
oscillates between 0 and $1/N$ on $S'^{1}$, which indicates that $c_{\textrm{1D}}$ is also nontrivial. 
Hence the NS with nontrivial $c_{\textrm{1D}}$ always accompanies a NL with nontrivial $\tilde{c}_{\textrm{1D}}$ \cite{SM}.

{\it Tight binding model.--}
\begin{figure}[t]
\includegraphics[width=\linewidth]{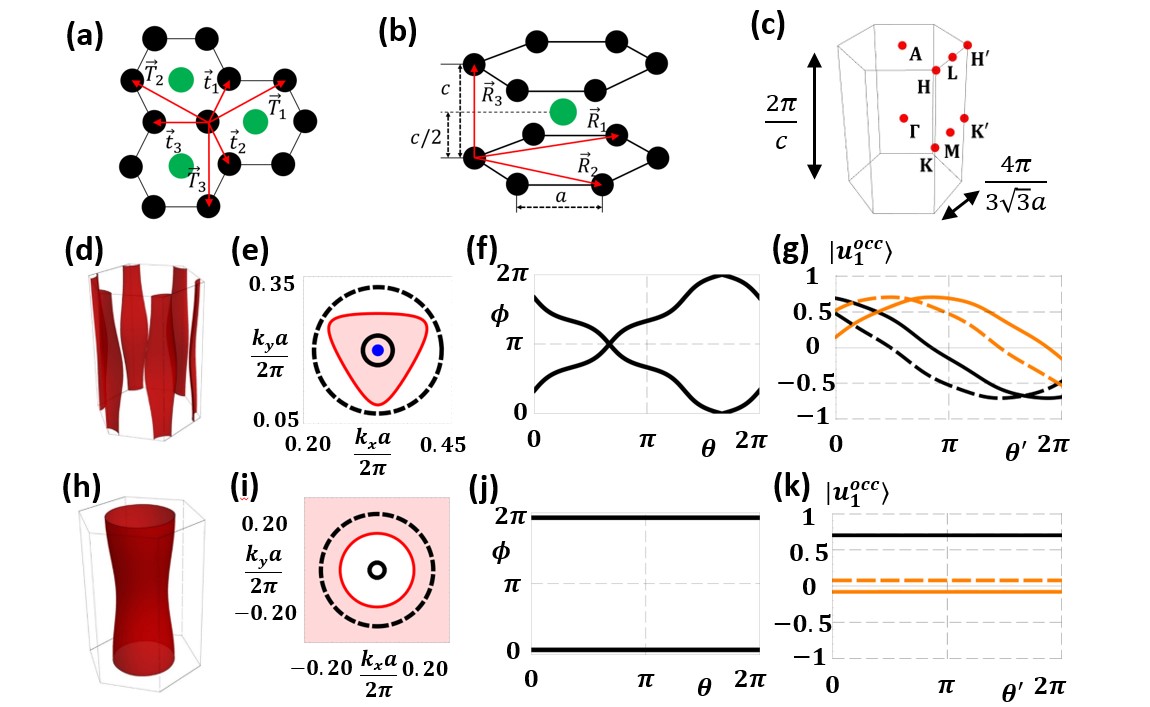}
\caption{
Lattice models on a hexagonal lattice composed of stacked honeycomb layers. 
(a) Structure of a honeycomb layer where a black dot indicates a $s$-orbtial. 
$\mathbf{t}_{1,2,3}$ and $\mathbf{T}_{1,2,3}$ are relative position vectors 
between nearest neighboring and next nearest neighboring sites, respectively. 
(b) A side view of the lattice. $\mathbf{R}_{1,2,3}$ denote the Bravais lattice vectors. 
The green dot indicates the extra lattice site added to construct 6-band models. 
(c) The first Brillouin zone (BZ) and its high symmetry points. 
(d) NSs (red) with $c_{\textrm{1D}}=1$ from the 4-band model with $E_{\mathrm{on}}=-0.5$, $t=-1$, $t_{z}=-0.1$, $\psi_{0}=0.5$. 
(e) A cross section of a NS on the $k_{z}=0$ plane where the blue dot at the center is the NL below $E_F$. 
The light red (white) region is where $\mathrm{det}[A(\mathbf{k})]$ is positive (negative). 
(f) The phase of the eigenvalues of $A_{FB}(\mathbf{k})$ along the black dashed line in (e).
(g) Components of $|u^{\mathrm{occ}}_{1\mathbf{k}}\rangle$ along the small black solid circle inside NS in (e). Black, orange solid (dashed)
curves correspond to the first, second (third, fourth) components of $|u^{\mathrm{occ}}_{1\mathbf{k}}\rangle$.
(h,i,j,k) correspond to the NS with $c_{\textrm{1D}}=0$ from the 4-band model with $E_{\mathrm{on}}=-2$, $t=-1$, $t_{z}=-0.1$, $\psi_{0}=0.5$.
}
\label{fig:BDI4band}
\end{figure}
We first construct a 4-band BdG Hamiltonian on a hexagonal lattice composed of vertically stacked honeycomb layers
shown in Fig.~\ref{fig:BDI4band}(a). 
A $s$-orbital is placed at each lattice site marked by black dots in Fig.~\ref{fig:BDI4band}(a). 
As the class BDI has full spin-rotation symmetry, we neglect the spin degrees of freedom.
The normal state is described by the Hamiltonian
\begin{align}
h_{2}(\mathbf{k}) &= \left(E_{\mathrm{on}}+2 t_{z} \cos(k_{z} c)\right)\mathbb{1}_{2\times 2}\nonumber \\
&+ t\sum_{i=1}^{3}\left(\cos (\mathbf{k}\cdot \mathbf{t}_{i})\tau_{x}+\sin (\mathbf{k}\cdot \mathbf{t}_{i})\tau_{y}\right),\label{hnor2}
\end{align}
where  $E_{\mathrm{on}}$ is the on-site energy, $t$ ($t_{z}$) is the intra-layer (inter-layer) nearest-neighbor (NN) hopping, 
$\tau_{i}$ are Pauli matrices for the sublattice degrees of freedom.
The Hamiltonian has inversion $\mathcal{I}$ and time-reversal $\mathcal{T}$ symmetries represented by $\mathcal{I}=\tau_{x}$ and $\mathcal{T}=\mathcal{K}$, respectively. 
We introduce an on-site odd-parity pairing function $\delta_{2}(\mathbf{k})=\psi_{0}\tau_{z}$ where $\psi_{0}$ is a real constant.
Then one can define a BdG Hamiltonian for spinless fermions (or a reduced BdG Hamiltonian) belonging to class BDI as 
$\mathcal{H}_{\mathrm{rBdG}}(\mathbf{k})=\begin{pmatrix}
h_{2}(\mathbf{k})&&\delta_2(\mathbf{k})\\
\delta^{*}_2(-\mathbf{k})&&-h^{T}_{2}(-\mathbf{k})
\end{pmatrix}$.

The NSs of $\mathcal{H}_{\mathrm{rBdG}}$ are shown in Fig.~\ref{fig:BDI4band}(d) where each NS encloses an edge of the first Brillouin zone (BZ) parallel to the $k_z$ axis. 
A cross section of a NS on the $k_{z}=0$ plane is plotted in Fig.~\ref{fig:BDI4band}(e) 
where the determinant of the off-diagonal block $A(\mathbf{k})$ of $\mathcal{H}_{\mathrm{rBdG}}(\mathbf{k})$ is positive (negative) in the red (white) region.
As $\mathrm{det}A(\mathbf{k})$ changes sign across the NS, $c_{\textrm{0D}}$ is non-trivial.

$c_{\textrm{1D}}$ is calculated on the dashed black line in Fig.~\ref{fig:BDI4band}(e).
As $\mathrm{det}A(\mathbf{k})<0$, the dashed black loop can be mapped to a loop in $\left[\mathrm{O}(2)-\mathrm{SO}(2)\right]$ by $A_{FB}(\mathbf{k})$.
$A_{FB}(\mathbf{k})$ can be restricted to $\mathrm{SO}(2)$ using a map $f : \left[\mathrm{O}(2)-\mathrm{SO}(2)\right]\rightarrow \mathrm{SO}(2)$ such that
$f(A)=\begin{pmatrix}
1&&0\\0&&-1
\end{pmatrix}A$, where $A\in\left[\mathrm{O}(2)-\mathrm{SO}(2)\right]$.
Since a $\mathrm{SO(2)}$ matrix has eigenvalues in the form of $\exp(\pm i\phi)$, $c_{\textrm{1D}}$ can be computed using the phase of $f(A_{FB}(\mathbf{k}))$ eigenvalues,
which is displayed  in Fig.~\ref{fig:BDI4band}(f) where the phase changes from $0$ to $2\pi$, hence $c_{\textrm{1D}}=1$.

To determine $\tilde{c}_{\textrm{1D}}$, in Fig.~\ref{fig:BDI4band}(g), we plot the components of $|u^{\mathrm{occ}}_{1}\rangle$
computed on a small black circle parametrized by $\theta'\in[0,2\pi]$ inside the NS shown in Fig.~\ref{fig:BDI4band}(e).
The opposite signs of $|u^{\mathrm{occ}}_{1}\rangle$ at $\theta'=0$ and $2\pi$ in Fig.~\ref{fig:BDI4band}(g) 
indicate $\tilde{c}_{\textrm{1D}}=1$ and the presence of a NL between occupied bands, marked by a blue dot in Fig.~\ref{fig:BDI4band}(e), which confirms
the linking structure of DCNSs.

As $E_{\mathrm{on}}$ decreases, the size of the NSs increases. At $E_{\mathrm{on}}\approx-0.67$, the NSs merge and form a single NS enclosing
the BZ center. The resulting NS has $c_{\textrm{0D}}=1$, $c_{\textrm{1D}}=\tilde{c}_{\textrm{1D}}=0$, and there is no NL inside the NS [see Fig.~\ref{fig:BDI4band}(h-k)].

It is straightforward to extend the above idea to general $2N$-band ($\mathrm{N}>2$) systems.
For instance, we can extend $\mathcal{H}_{\mathrm{rBdG}}$ to a 6-band model by adding an extra s-orbital at the center between two hexagons in adjacent honeycomb layers
as in Fig.~\ref{fig:BDI4band}(b). 
The corresponding $c_{\textrm{0D}}$ and $\tilde{c}_{\textrm{1D}}$ can be detemined by using the same way as above.
To evaluate $c_{\textrm{1D}}$,  as there are three occupied bands,
$A_{FB}$ becomes a $\mathrm{SO(3)}$ matrix.
Thus the homotopy equivalence class of a closed loop in SO(3) should be determined.
Further generalization to $2N$-band systems is straightforward \cite{SM}.

\begin{figure}[t]
\centering
\includegraphics[width=8.5cm]{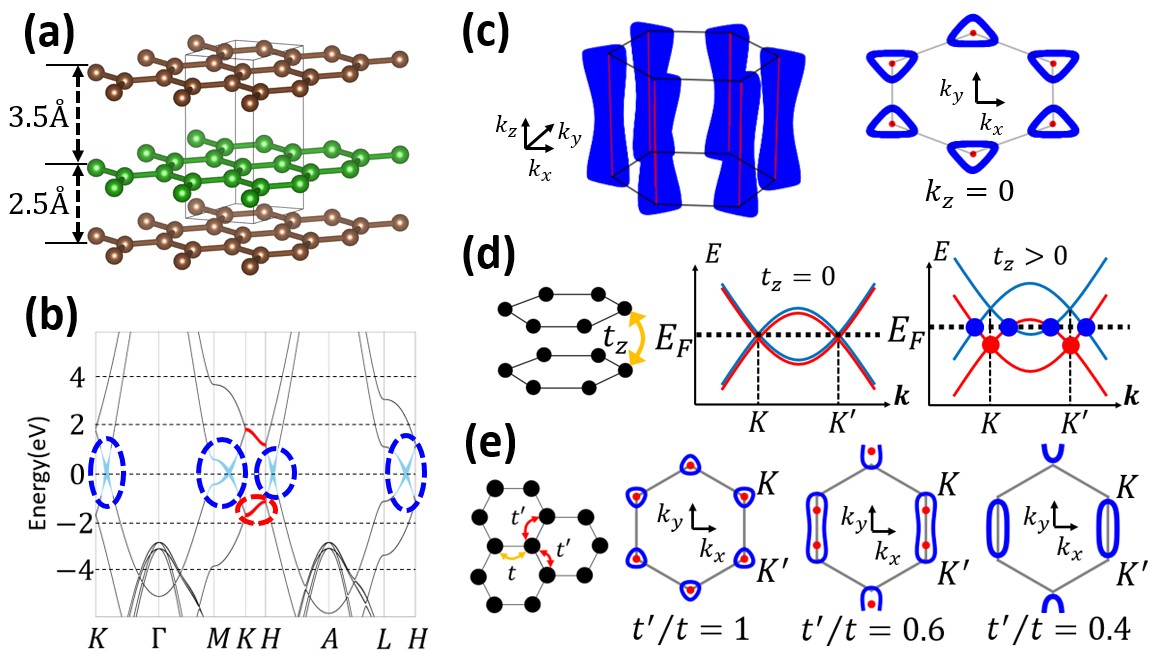}
\caption{
(a) Atomic structure of AA-stacked bilayer graphene multilayers (ABGM). 
(b) The relevant band structure from first-principles calculations. The band crossing at (below) $E_F$ is emphasized by blue (red) circles. 
(c) Nodal structure in the first BZ and the cross section in the $k_z=0$ plane. The blue surfaces (red lines) are NSs at $E_F$ (NLs below $E_F$). 
(d) The origin of nodal structures in (c). 
The interlayer hopping $t_z$ splits the band structure of two graphene layers hosting Dirac points at BZ corners. The band crossing at $E_F$ results in a NL (blue dots) enclosing a Dirac point (red dots) below $E_F$.
(e) Topological phase transition induced by strain represented by anisotropic hopping $t\neq t'$.
Merging of two NSs (blue) followed by pair-annihilation of NLs (red) gives NSs with $c_{\textrm{1D}}=0$. 
}
\label{fig:DFT}
\end{figure}

{\it DCNSs in AA-stacked bilayer graphene multilayers (ABGM).---}  
DCNSs can appear not only in nodal superconductors but also in semimetals,
because spinless fermion systems with inversion and sublattice symmetries can also belong to class BDI~\cite{bzduvsek2017doubly}.
Motivated by the model proposal in Ref.~\cite{bzduvsek2017doubly}, we performed first-principles calculations of ABGM [see Fig.~\ref{fig:DFT}(a)].
As shown in Fig.~\ref{fig:DFT}(b,c), the system has DCNSs at $E_F$ enclosing NLs below $E_F$.
The nodal structure of ABGM can easily be understood from that of bilayer graphene in Fig.~\ref{fig:DFT}(d).
Each graphene has two Dirac points (DPs) at $E_F$, protected by $\mathfrak{T}$ symmetry.
The nonzero interlayer hopping $t_z$ splits the degenerate band structure of graphene bilayer such that
the band crossing at $E_F$ generates NLs, which naturally enclose DPs below $E_F$.
Simple extension of this 2D band structure along the $k_z$-direction gives the DCNSs in Fig.~\ref{fig:DFT}(c).
This example clearly demonstrates that vertical stacking of 2D Dirac semimetal bilayers generally hosts DCNSs as long as the sublattice and inversion symmetries are protected.
We note that although the chiral symmetry of ABGM is not an exact symmetry, as the low energy band structure including the NSs and NLs near $E_F$ has effective chiral symmetry,
the linked nodal structure we propose can be observed in this system.

There are other material proposals of class BDI semimetals via lateral stacking of 1D semimetals~\cite{wu2018nodalsurface,zhong2016BDImat1,chen2020BDImat2}, which we found to host trivial NSs.
However, even in this case, we propose that trivial NSs can turn into DCNSs by applying strain. For instance, by applying in-plane strain,  
DCNSs of ABGM can be transformed to trivial NSs [see Fig.~\ref{fig:DFT}(e)], which is confirmed by a tight-binding model for ABGM~(see also \cite{pereira2009tight,cocco2010gap,naumov2011gap}).
Such a topological phase transition between DCNSs and trivial NSs can generally occur in class BDI semimetals through the mechanism called {\it double band inversion} \cite{SM}.
This means that trivial NSs in proposed materials can also be transformed to DCNSs under suitable perturbations through the double band inversion process.

{\it Discussion.---}
DCNSs in class D superconductors also exhibit linking structure with NPs below $E_F$~\cite{bzduvsek2017doubly}.
As shown in ~\cite{bzduvsek2017doubly}, the 2D charge $c_{\mathrm{2D}}$ ($\tilde{c}_{\mathrm{2D}}$) of the NS (NP)
satisfies $c_{\mathrm{2D}}=-2\tilde{c}_{\mathrm{2D}}$. 
As $\tilde{c}_{\mathrm{2D}}$ can be nonzero when NPs exist below $E_F$, inside the NS, 
a NS with nonzero $c_{\mathrm{2D}}$ always accompanies NPs inside it, demonstrating the linking structure in Fig.~\ref{fig:geolink}(c). 

To conclude, we have established the linking structure of DCNSs in class BDI superconductors and semimetals.
Combining the related works on class AI~\cite{ahn2018linking,bzduvsek2019linking}, CI~\cite{ahn2018linking,bzduvsek2019linking}, D~\cite{bzduvsek2017doubly}, 
we have completed the fundamental relation between the doubly charged nodes and their linking structures  based on the $\mathrm{AZ}+\mathcal{I}$ classification.
As there are various doubly charged nodal structures in systems with dimension $d>3$~\cite{lian2016five},
investigating possible linking structures of higher-dimensional gapless topological states is
an interesting direction for future study. 

\begingroup
\renewcommand{\addcontentsline}[3]{}
\renewcommand{\section}[2]{}
\begin{acknowledgments}
We thank J. Ahn for useful comments.
S.K. and B.-J.Y. were supported by the Institute for Basic Science in Korea (Grant No. IBS-R009-D1), 
Samsung  Science and Technology Foundation under Project Number SSTF-BA2002-06,
the National Research Foundation of Korea (NRF) grant funded by the Korea government (MSIT) (No.2021R1A2C4002773), 
and the U.S. Army Research Office and Asian Office of Aerospace Research \& Development (AOARD) under Grant No. W911NF-18-1-0137.
D.-C.R. was supported by the Institute for Basic Science in Korea (Grant No. IBS-R009-D1), 
the National Research Foundation of Korea (NRF) grant funded by the Korea government (MSIT) (No.2021R1A2C4002773).
\end{acknowledgments}

\endgroup

\clearpage

\renewcommand{\appendixpagename}{\center \Large Supplemental Material (SM)}

\appendixpage
\setcounter{page}{1}
\setcounter{section}{0}
\setcounter{figure}{0}
\setcounter{equation}{0}
\renewcommand{\thefigure}{S\arabic{figure}}
\renewcommand{\theequation}{S\arabic{equation}}
\renewcommand{\thesection}{S\arabic{section}}
\tableofcontents
\hfill

\section{Class BDI}
\begin{figure}[b]
\includegraphics[width=\linewidth]{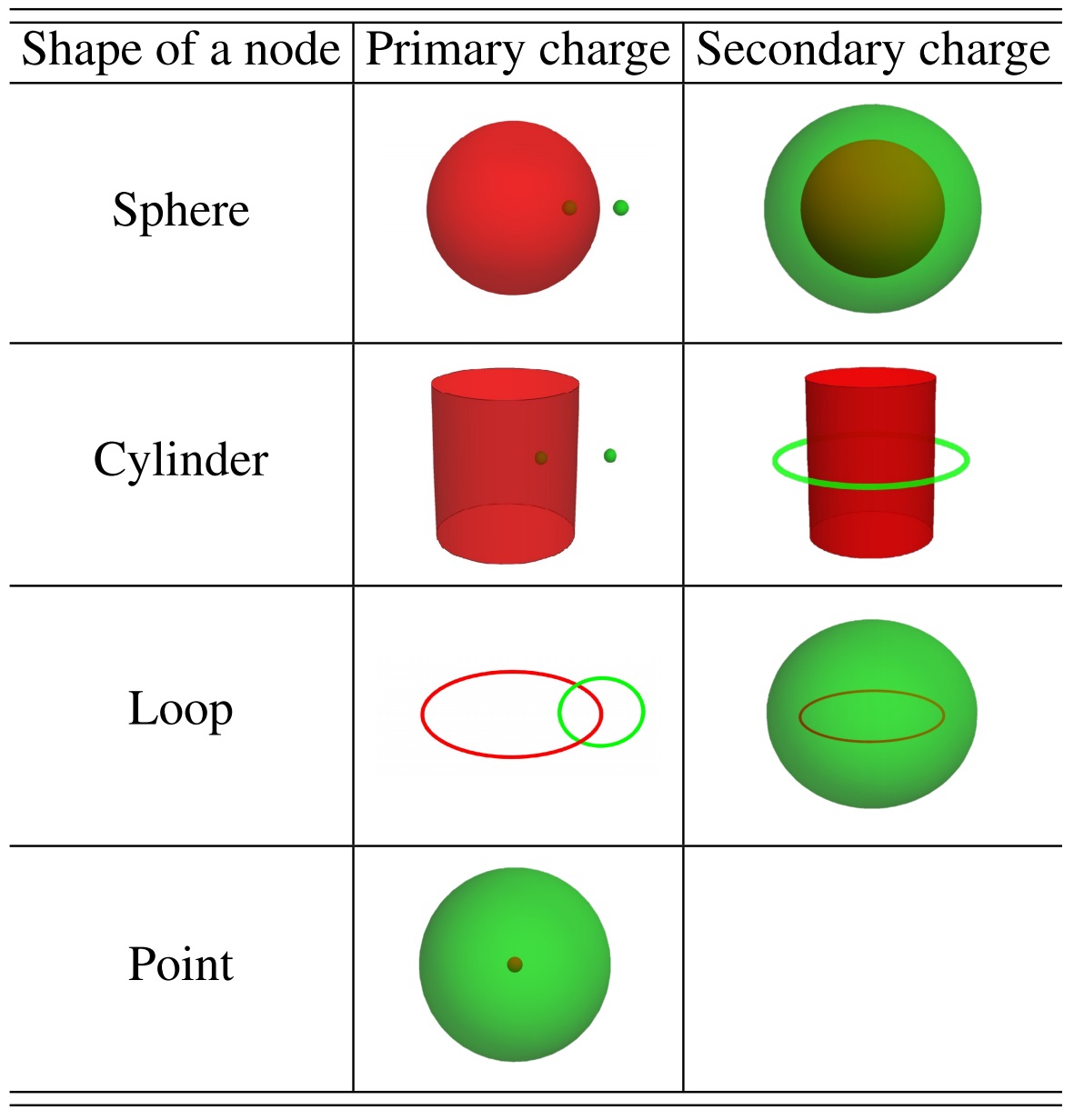}
\caption{
Schematic description of the manifold enclosing a node on which the primary and the secondary topological charges are defined.
A nodal structure (an enclosing manifold) is depicted in red (green).
}
\label{fig:nodeshp&TC}
\end{figure}
\subsection{Topological charges of class BDI}

Let us first consider the nodal surface which generally has a cylindrical shape
shown in Fig.~\ref{fig:nodeshp&TC}. The nodal surface has two types of topological charges: a 0D charge and a 1D charge.

 The 0D charge can be defined as follows.
 Due to the symmetry $\mathcal{C}=\sigma_z$, the Hamiltonian $H(\mathbf{k})$, which is a $2N\times 2N$ matrix, takes a block off-diagonal form as 
\begin{align}
H(\mathbf{k})=\begin{pmatrix} 0&A(\mathbf{k})\\A^{T}(\mathbf{k})&0 \end{pmatrix},
\end{align}
where $A(\mathbf{k})$ denotes an $N\times N$ real matrix.
 The 0D charge $c_{\mathrm{BDI}}(S^{0})$ is defined using $A(\mathbf{k})$ as
\begin{align}
c_{\mathrm{BDI}}(S^{0})=\mathrm{sign}\{\mathrm{det}A(\mathbf{k}_{\text{in}})\cdot \mathrm{det}A(\mathbf{k}_{\text{out}})\},
\label{0D_BDI}
\end{align}
where $S^{0}=\{\mathbf{k}_{\text{in}},\mathbf{k}_{\text{out}}\}$ and $\mathbf{k}_{\text{in}}$ ($\mathbf{k}_{\text{out}}$) indicates a momentum inside (outside) the nodal surface~\cite{bzduvsek2017doubly}.

The definition of the 0D charge can be understood in terms of a band inversion process across the nodal surface.
 Suppose there are $N$ occupied bands and $N$ unoccupied bands, and the energies of unoccupied bands $\{E_{n\mathbf{k}}\}$~$(n=1,\cdots,N)$ are aligned as $0\leq E_{1\mathbf{k}} \leq \cdots \leq E_{N\mathbf{k}}$.
 Since $\{H(\mathbf{k}), \mathcal{C}\}=0$, for each occupied state $|u_{n\mathbf{k}}^{\text{occ}}\rangle$ with the energy $-E_{n\mathbf{k}}$,
there is a relevant unoccupied state $|u_{n\mathbf{k}}^{\text{unocc}}\rangle\propto \mathcal{C} |u_{n\mathbf{k}}^{\text{occ}}\rangle$ with the energy $E_{n\mathbf{k}}$.
 Also $|u_{n\mathbf{k}}^{\text{occ}}\rangle$ and $|u_{n\mathbf{k}}^{\text{unocc}}\rangle$ can be chosen as
\begin{gather}
|u_{n\mathbf{k}}^{\text{occ}}\rangle ={1 \over \sqrt{2}} \begin{pmatrix}
|u_{n\mathbf{k}}^{\uparrow}\rangle \\ |u_{n\mathbf{k}}^{\downarrow}\rangle
\end{pmatrix},\; 
|u_{n\mathbf{k}}^{\text{unocc}}\rangle ={1 \over \sqrt{2}} \begin{pmatrix}
|u_{n\mathbf{k}}^{\uparrow}\rangle \\ -|u_{n\mathbf{k}}^{\downarrow}\rangle
\end{pmatrix},
\label{BDIstate}
\end{gather}
where $|u_{n\mathbf{k}}^{\uparrow}\rangle,~ |u_{n\mathbf{k}}^{\downarrow}\rangle(n=1,\cdots ,N)$ are $N$-dimensional vectors that satisfy $\langle u^{\uparrow}_{n\mathbf{k}}|u^{\uparrow}_{m\mathbf{k}} \rangle=\langle u^{\downarrow}_{n\mathbf{k}}|u^{\downarrow}_{m\mathbf{k}} \rangle=\delta_{nm}$.
 Using these $N$-dimensional vectors, $A(\mathbf{k})$ can be expressed as
\begin{gather}
A(\mathbf{k})=\sum_{n=1}^{N} E_{n\mathbf{k}}|u^{\uparrow}_{n\mathbf{k}}\rangle\langle u^{\downarrow}_{n\mathbf{k}}|.
\label{A}
\end{gather}

Suppose that, at one side of the nodal surface, the highest occupied state $|u^{\mathrm{occ}}_{1\mathbf{k}}\rangle$ and the lowest unoccupied state $|u^{\mathrm{unocc}}_{1\mathbf{k}}\rangle$ are given by Eq.~(\ref{BDIstate}).
 If there is a band inversion across the nodal surface, $|u^{\mathrm{occ}}_{1\mathbf{k}}\rangle$ and $|u^{\mathrm{unocc}}_{1\mathbf{k}}\rangle$ at the other side of the nodal surface are given by
\begin{gather}
|u_{1\mathbf{k}}^{\text{occ}}\rangle ={1 \over \sqrt{2}} \begin{pmatrix}
|u_{1\mathbf{k}}^{\uparrow}\rangle \\ -|u_{1\mathbf{k}}^{\downarrow}\rangle
\end{pmatrix},\; 
|u_{1\mathbf{k}}^{\text{unocc}}\rangle ={1 \over \sqrt{2}} \begin{pmatrix}
|u_{1\mathbf{k}}^{\uparrow}\rangle \\ |u_{1\mathbf{k}}^{\downarrow}\rangle
\end{pmatrix}.\label{InvBDIstate}
\end{gather}
After the band inversion, $A(\mathbf{k})$ changes from Eq.~(\ref{A}) to
\begin{gather}
A(\mathbf{k})=-E_{1\mathbf{k}}|u^{\uparrow}_{1\mathbf{k}}\rangle\langle u^{\downarrow}_{1\mathbf{k}}|+\sum_{n=2}^{N} E_{n\mathbf{k}}|u^{\uparrow}_{n\mathbf{k}}\rangle\langle u^{\downarrow}_{n\mathbf{k}}|.
\label{A'}
\end{gather} 
Since each of $\left\{|u^{\uparrow}_{n\mathbf{k}}\rangle\right\}$ and $\left\{|u^{\downarrow}_{n\mathbf{k}}\rangle\right\}$ satisfies the orthonormality condition, $\mathrm{det}A(\mathbf{k})$ for Eq.~(\ref{A}),~(\ref{A'}) should be either $E_{1}\cdots E_{N}$ or $-E_{1}\cdots E_{N}$.
 The signs of the determinants are determined by relative orientations between the bases $\left\{|u^{\uparrow}_{n\mathbf{k}}\rangle\right\}$ and $\left\{|u^{\downarrow}_{n\mathbf{k}}\rangle\right\}$.
 For example, for Eq. (\ref{A}), the determinant of $A(\mathbf{k})$ is $E_{1\mathbf{k}}\cdots E_{N\mathbf{k}}$ when $\left\{|u^{\uparrow}_{n\mathbf{k}}\rangle\right\}$ and $\left\{|u^{\downarrow}_{n\mathbf{k}}\rangle\right\}$ have the same orientation.
  Regardless of the relative orientation between these bases, however, $\mathrm{det}A(\mathbf{k})$s for Eq.~(\ref{A}) and (\ref{A'}) have the opposite signs.
 Therefore $c_{\mathrm{BDI}}(S^{0})=-1$ when there is a band inversion across the nodal surface.
 
In fact, the band inversion between the highest occupied band and the lowest unoccupied band is the only allowed change of the eigenstates across the nodal surface.
 At the nodal surface, both the energy of the highest occupied state and that of the lowest unoccupied state are zero.
 This means that the highest occupied state and lowest unoccupied state can be discontinuous across the nodal surface.
 On the other hand, the other states change continuously across the nodal surface because their energies are generally non-degenerate at the nodal surface.
 Then the highest occupied and lowest unoccupied states at one side of the nodal surface is given by linear combinations of them at the other side of the nodal surface. 
 But not all linear combinations are possible as $\mathfrak{T}$ and $\mathcal{C}$ symmetries have to be satisfied. 
 Considering these symmetries, we can find that a band inversion is the only possible change of the eigenstates across the nodal surface.

Now let us consider the 1D charge of the nodal surface. For this, we first consider the spectral flattening of the Hamiltonian by smoothly deforming the band structure so that all the energies $E_{1\mathbf{k}},\cdots,E_{N\mathbf{k}}$ become 1. After the flattening, $A(\mathbf{k})$ in Eq.~(\ref{A}) becomes an element of $\mathrm{O(N)}$.
 Then the 1D charge $c_{\mathrm{BDI}}(S^{1})$ can be defined as
\begin{gather}
c_{\mathrm{BDI}}(S^{1})=\left[A_{FB}:S^{1}\rightarrow\mathrm{O(N)}\right].
\end{gather}
Here $A_{FB}$ is an off-diagonal block of the flattened Hamiltonian and $S^{1}$ is a circle encircling the nodal surface [see Fig.~\ref{fig:nodeshp&TC}].
 $\left[A_{FB}:S^{1}\rightarrow\mathrm{O(N)}\right]$ means the homotopy equivalence class within the $N$-dimensional orthogonal group.
 
In the case of the nodal line below the Fermi level, there are two different ways of describing its 1D charge.
 One is to take the representation $\mathfrak{T}=\mathcal{K}$ so that the eigenstates become real-valued.
 For a nodal line formed between two occupied bands $|u_{n,\mathbf{k}}^{\mathrm{occ}}\rangle$ and $|u_{n+1,\mathbf{k}}^{\mathrm{occ}}\rangle$,
 its topological charge can be defined on a circle, enclosing the nodal line, which is parametrized by an angle $\theta\in[-\pi,\pi]$.
 If we assume that the state $|u_{i,\mathbf{k}}^{\mathrm{occ}}\rangle$ $(i=n, n+1)$ changes continuously for $\theta\in(-\pi,\pi)$,
 the following relation should be satisfied at $\theta=\pm \pi$~\cite{ahn2018linking},
\begin{gather}
|u_{i,\pi}^{\mathrm{occ}}\rangle=\pm|u_{i,-\pi}^{\mathrm{occ}}\rangle.
\label{1stSW}
\end{gather}
 When the state $|u_{i,\mathbf{k}}^{\mathrm{occ}}\rangle$ changes discontinuously (continuously) at $\theta=\pm \pi$,  
 $|u_{i,\mathbf{k}}^{\mathrm{occ}}\rangle$ does (does not) undergo an orientation-reversal on $S^{1}$, which indicates the nontrivial (trivial)
 1D topological charge of the nodal line~\cite{ahn2018linking}.
 
The second way is to choose a smooth complex gauge and compute the winding number of the eigenstates.
 Consider an effective Hamiltonian of two occupied bands $|u^{\mathrm{occ}}_{n,\mathbf{k}}\rangle$ and $|u^{\mathrm{occ}}_{n+1,\mathbf{k}}\rangle$,
\begin{gather}
H_{\textrm{eff}}(\mathbf{k})=|u_{n,\mathbf{k}}^{\textrm{occ}}\rangle\langle u_{n,\mathbf{k}}^{\textrm{occ}}|-|u_{n+1,\mathbf{k}}^{\textrm{occ}}\rangle\langle u_{n+1,\mathbf{k}}^{\textrm{occ}}|.
\end{gather}
 Since the eigenvalues of this Hamiltonian are $\pm 1$, there can be an effective chiral symmetry $\mathcal{C}_{\mathrm{eff}}$ so that $\left\{H_{\mathrm{eff}}(\mathbf{k}), \mathcal{C}_{\mathrm{eff}}\right\}=0$.
 One can show that the effective Hamiltonian of $4$-band model has an effective chiral symmetry.
 Focusing on the $4$-band model, we can get the off-diagonal block $A_{\mathrm{eff}}(\mathbf{k})$ of the effective Hamiltonian after appropriate basis transformation.
 The 1D charge of the nodal line can be defined using $A_{\mathrm{eff}}(\mathbf{k})$ by
\begin{align}
\tilde{c}_{\textrm{1D}}(\tilde{S^{1}})&=\frac{i}{2\pi}\oint_{\tilde{S^{1}}} d\mathbf{k}\cdot \mathrm{tr}\left[(A_{\textrm{eff}}(\mathbf{k}))^{\dagger} \nabla A_{\textrm{eff}}(\mathbf{k})\right],
\end{align}
where $\tilde{S}^{1}$ is a circle encircling the nodal line between two occupied bands $|u^{\mathrm{occ}}_{n,\mathbf{k}}\rangle$ and $|u^{\mathrm{occ}}_{n+1,\mathbf{k}}\rangle$.
 In Sec III.B, we show that $\tilde{c}_{\textrm{1D}}$ is quantized in the $4$-band model.

\subsection{Linking structure of class BDI}

\subsubsection{Case of 4 bands}
Let us consider a 4-band model with energies $\pm E_{1}$, $\pm E_{2}$ ($0\leq E_{1}\leq E_{2}$). 
Corresponding $A(\mathbf{k})$ can be parametrized by two angles $\theta(\mathbf{k})$ and $\phi(\mathbf{k})$ as
\begin{align}
A_\pm(\mathbf{k})&=\frac{E_{2}\mp E_{1}}{2}\begin{pmatrix}
\sin\theta(\mathbf{k}) & \cos\theta(\mathbf{k}) \\ \cos\theta(\mathbf{k}) & -\sin\theta(\mathbf{k})
\end{pmatrix}\nonumber\\
&+\frac{E_{2}\pm E_{1}}{2} \begin{pmatrix}
\cos\phi(\mathbf{k}) & -\sin\phi(\mathbf{k}) \\ \sin\phi(\mathbf{k}) & \cos\phi(\mathbf{k})
\end{pmatrix}\label{Apm},
\end{align}
where $\mathrm{det}A_\pm(\mathbf{k})=\pm E_1E_2$.
 We note that two Hamiltonians described by $A_{+}(\mathbf{k})$ and $A_-(\mathbf{k})$, respectively, are related by a band inversion between $|u_{1\mathbf{k}}^{\text{occ}}\rangle$ and $|u_{1\mathbf{k}}^{\text{unocc}}\rangle$, and the corresponding band crossing points form a nodal surface.
 This is consistent with the fact that the 0D charge of the nodal surface is given by Eq. (\ref{0D_BDI}).
 
 To determine the 1D charge of the the nodal surface, we assume that the Hamiltonian outside (inside) the nodal surface is described by $A_-(\mathbf{k})\left(A_+(\mathbf{k})\right)$.
 After flattening the Hamiltonian, $A_-(\mathbf{k})$ depends on $\theta(\mathbf{k})$ only. 
 Then $c_{\textrm{BDI}}(S^{1})$ is given by 
\begin{gather}
c_{\textrm{BDI}}(S^{1})=\frac{1}{2\pi}\oint_{S^{1}} d\mathbf{k}\cdot \nabla \theta(\mathbf{k})
\end{gather}
where $S^{1}$ is a circle surrounding the nodal surface.

Now let us show that the 1D charge $c_{\textrm{BDI}}(S^{1})$ of the nodal surface is identical to the 1D charge $\tilde{c}_{1\mathrm{D}}$ of a nodal line formed between occupied bands, which is inside the nodal surface. Since $A_+(\mathbf{k})$ is an off-diagonal block of the Hamiltonian defined inside the nodal surface, $\tilde{c}_{1\mathrm{D}}$ can be determined by $A_+(\mathbf{k})$ and the corresponding occupied states $|u_{1,2\mathbf{k}}^{\text{occ}}\rangle$.
 The winding number of the nodal line formed between occupied bands can be evaluated using an effective two-band Hamiltonian given by
\begin{gather}\label{eq:Heff}
H_{\textrm{eff}}(\mathbf{k})=|u_{1\mathbf{k}}^{\textrm{occ}}\rangle\langle u_{1\mathbf{k}}^{\textrm{occ}}|-|u_{2\mathbf{k}}^{\textrm{occ}}\rangle\langle u_{2\mathbf{k}}^{\textrm{occ}}|.
\end{gather}
Plugging the explicit form of $|u_{1,2\mathbf{k}}^{\text{occ}}\rangle$ into Eq.~(\ref{eq:Heff}), we find that $H_{\textrm{eff}}(\mathbf{k})$, expressed in terms of $\theta$ and $\phi$, has an effective chiral symmetry so that it can be transformed to a block off-diagonal form with the off-diagonal block $A_{\textrm{eff}}(\mathbf{k})$ given by
\begin{gather}
A_{\textrm{eff}}(\mathbf{k})=\frac{1}{2}i e^{-i\theta(\mathbf{k})} \begin{pmatrix}
-e^{i\phi(\mathbf{k})} & 1\\1&-e^{-i\phi(\mathbf{k})}\label{Aeff}
\end{pmatrix}.
\end{gather}
 In terms of $A^{\textrm{eff}}(\mathbf{k})$, $\tilde{c}_{\textrm{1D}}$ is given by
\begin{align}
\tilde{c}_{\textrm{1D}}(\tilde{S^{1}})&=\frac{i}{2\pi}\oint_{\tilde{S^{1}}} d\mathbf{k}\cdot \mathrm{tr}\left[(A^{\textrm{eff}}(\mathbf{k}))^{\dagger} \nabla A^{\textrm{eff}}(\mathbf{k})\right]\nonumber\\
&=\frac{1}{2\pi}\oint_{\tilde{S^{1}}} d\mathbf{k}\cdot \nabla \theta(\mathbf{k}),
\label{BDI4BNLch}
\end{align}
where $\tilde{S^{1}}$ is a circle surrounding the nodal line inside the nodal surface.
 Since $\theta(\mathbf{k})$ is continuously defined across the nodal surface, $\tilde{c}_{\textrm{1D}}(\tilde{S^{1}})$ and $c_{\mathrm{BDI}}(S^{1})$ are the same.

 To confirm that the doubly charged nature of the nodal surface is originated from its linking structure with nodal lines between occupied bands, we have to show that the nontrivial $\tilde{c}_{1D}(\tilde{S}^{1})$ arises from the nodal line between the occupied bands.
 At each nodal line, $A_{+}(\mathbf{k})$ is $\theta$-independent because $E_{1}=E_{2}$.
 This means that, inside the nodal surface, $\theta$ can have non-trivial winding around the nodal line.
 On the other hand, $\theta$ cannot have non-trivial winding inside the nodal surface when there isn't any nodal lines because the Hamiltonian always has well-defined $\theta$ dependent term inside the nodal surface.
 Therefore, $\tilde{c}_{1D}(\tilde{S}^{1})$ can be non-trivial only if $\tilde{S}^{1}$ surrounds a nodal line inside the nodal surface.
 
\subsubsection{Cases of $2N$ bands ($N>2$)}
Now we consider general the cases of $2N$ bands with $N>2$.
 On both sides of the nodal surface, the possible forms of $A(\mathbf{k})$ are described by Eq.~(\ref{A}) and (\ref{A'}).
 After flattening the energy spectrum, the corresponding off-diagonal blocks of a flat-band Hamiltonian with $2N$ bands are given by
\begin{gather}
A^{\pm}_{FB}(\mathbf{k})=\pm |u^{\uparrow}_{1\mathbf{k}}\rangle\langle u^{\downarrow}_{1\mathbf{k}}|+\sum_{n=2}^{N} |u^{\uparrow}_{n\mathbf{k}}\rangle\langle u^{\downarrow}_{n\mathbf{k}}|,
\end{gather}
where $A^{+}_{FB}(\mathbf{k})$ ($A^{-}_{FB}(\mathbf{k})$) corresponds to the Hamiltonian defined inside (outside) of the nodal surface.
 It is obvious that $A_{FB}(\mathbf{k})$ changes discontinuously across the nodal surface.
 To describe the topological charge, we additionally introduce $A'_{FB}(\mathbf{k})$ given by 
\begin{gather}\label{eq:A_prime}
A'_{FB}(\mathbf{k})=-|u^{\uparrow}_{1\mathbf{k}}\rangle\langle u^{\downarrow}_{1\mathbf{k}}|+\sum_{n=2}^{N} |u^{\uparrow}_{n\mathbf{k}}\rangle\langle u^{\downarrow}_{n\mathbf{k}}|,
\end{gather}
which is defined inside the nodal surface. 
 $A'_{FB}(\mathbf{k})$ and $A^{-}_{FB}(\mathbf{k})$ have the same form but are defined in different regions of the momentum space, i.e., inside and outside the nodal surface, respectively.

 To evaluate the 1D charge of the nodal surface, let us consider a circle $S^{1}$ surrounding it.
 The 1D charge is given by the homotopy equivalence class of $A^{-}_{FB}(\mathbf{k})$ defined on the circle $S^{1}$.
 As noted above, $A^{-}_{FB}(\mathbf{k})$ defined outside the nodal surface is continuously connected with $A'_{FB}$ defined inside the nodal surface.
 Hence, the 1D charge defined in terms of $A^{-}_{FB}(\mathbf{k})$ outside the nodal surface can be equivalently described by $A'_{FB}(\mathbf{k})$ defined inside the nodal surface as follows: 
\begin{gather}
c_{\mathrm{BDI}}(S^{1})=\left[A'_{FB}:S'^{1}\rightarrow \mathrm{O(N)}\right],
\end{gather}
where $S'^{1}$ is a circle inside the nodal surface, which is obtained by deforming $S^{1}$ continuously.
 
 Now we ask whether $S'^{1}$ can be shrunk to a point while keeping $A'_{FB}$ well-defined on it. In general, such a smooth deformation is impossible when there is a nodal line inside $S'^{1}$ at the energy satisfying $-E_{1}=-E_{2}$. This is because $|u^{\mathrm{occ}}_{1\mathbf{k}}\rangle$ and $|u^{\mathrm{occ}}_{2\mathbf{k}}\rangle$ cannot be uniquely specified at the nodal line due to the degeneracy so that $A'_{FB}$ cannot be defined as well [see Eq.~(\ref{eq:A_prime})]. Let us note that the presence of other nodal lines at the energy $-E_{n}=-E_{n+1}$ ($n>1$) does not affect $A'_{FB}(\mathbf{k})$.

Let us shrink $S'^{1}$ encircling the nodal line at the energy satisfying $-E_{1}=-E_{2}$, and see how $A'_{FB}(\mathbf{k})$ changes on $S'^{1}$. There are two possible behaviors of $A'_{FB}(\mathbf{k})$ expected on $S'^{1}$ : staying nearly constant or oscillating prominently along $S'^{1}$.
 In the former case, the homotopy equivalence class of $A'_{FB}$ defined on $S'^{1}$ should be trivial, so that the 1D charge of the nodal surface is trivial.
 On the other hand, in the latter case, the homotopy equivalence class should be non-trivial.
 Therefore, the 1D charge of the nodal surface is non-trivial.

In short, whether the 1D charge of the nodal surface is trivial or not can be determined from the behavior of $A'_{FB}(\mathbf{k})$ on a small circle encircling the nodal line at the energy satisfying $-E_{1}=-E_{2}$ inside the nodal surface.
 Interestingly, this information is sufficient to characterize the 1D charge of the nodal surface because the fundamental group of the orthogonal group is given by
\begin{gather}
\pi_{1}(\mathrm{O(N)})=\mathbb{Z}_{2},
\end{gather}
when $\mathrm{N}>2$.

The off-diagonal block $A^{+}_{FB}(\mathbf{k})$ of the flattened Hamiltonian is well-defined inside the nodal surface.
 Therefore, $A^{+}_{FB}(\mathbf{k})$ should be nearly constant along $S'^{1}$ when $S'^{1}$ is sufficiently close to the nodal line.
 As a result, one can consider the homotopy equivalence class of $A^{+}_{FB}(\mathbf{k})-A'_{FB}(\mathbf{k})$, instead of that of $A'_{FB}(\mathbf{k})$, to determine the 1D charge of the nodal surface.
 Since $A^{+}_{FB}(\mathbf{k})-A'_{FB}(\mathbf{k})$ is given by
\begin{gather}
A_{FB}(\mathbf{k})-A'_{FB}(\mathbf{k})=2|u^{\uparrow}_{1\mathbf{k}}\rangle\langle u^{\downarrow}_{1\mathbf{k}}|,
\end{gather}
the behavior of $|u^{\uparrow}_{1\mathbf{k}}\rangle$ and $|u^{\downarrow}_{1\mathbf{k}}\rangle$ along $S'^{1}$ should determine the 1D charge of the nodal surface.

If the circle $S'^{1}$ encircles a nodal line between the topmost and second topmost occupied bands, then $|u^{\mathrm{occ}}_{1\mathbf{k}}\rangle$ undergoes an orientation-reversal on $S'^{1}$.
 Therefore, $|u^{\uparrow}_{1\theta}\rangle$ and $|u^{\downarrow}_{1\theta}\rangle$ also undergo orientation-reversals [see Eq.~(\ref{BDIstate})].
 Here $\theta$ denotes the angle parametrizing $S'^{1}$.
 From $\langle u^{\uparrow(\downarrow)}_{1\theta}|u^{\uparrow(\downarrow)}_{1\theta}\rangle=1$, it is easy to show that, for some $i_{0}$ and $j_{0}$, the $i_{0}$-th component of $|u^{\uparrow}_{1\pi}\rangle$ ($\left[|u_{1\pi}^{\uparrow}\rangle\right]_{i_0}$) and the $j_{0}$-th component of $|u^{\downarrow}_{1\pi}\rangle$ ($\left[|u_{1\pi}^{\downarrow}\rangle\right]_{j_0}$) satisfy
\begin{gather}
\left|\left[|u_{1\pi}^{\uparrow}\rangle\right]_{i_0}\right|,\left|\left[|u_{1\pi}^{\downarrow}\rangle\right]_{j_0}\right| \geq \frac{1}{\sqrt{N}}.
\label{revcomp}
\end{gather}
Since $|u^{\uparrow}_{1\theta}\rangle$ and $|u^{\downarrow}_{1\theta}\rangle$ undergo orientation-reversals along $S'^{1}$, $\left[|u_{1\theta}^{\uparrow}\rangle\right]_{i_0}$ and $\left[|u_{1\theta}^{\downarrow}\rangle\right]_{j_0}$ cross 0 at some $\theta$.

Now let us consider an $N\times N$ matrix $|u^{\uparrow}_{1\theta}\rangle\langle u^{\downarrow}_{1\theta}|$ which is proportional to $A^{+}_{FB}(\mathbf{k})-A'_{FB}(\mathbf{k})$.
 From Eq.~(\ref{1stSW}) and (\ref{revcomp}), one can see that the $(i_{0},j_{0})$-component of $|u^{\uparrow}_{1\theta}\rangle\langle u^{\downarrow}_{1\theta}|$ oscillates between 0 and $1/N$ along $S'^{1}$.
 This corresponds to the case when the nodal surface carries a non-trivial 1D charge.

\subsubsection{Double band inversion : topological phase transition between DCNSs and trivial NSs}
\begin{figure}[t]
\centering
\includegraphics[width=8.5cm]{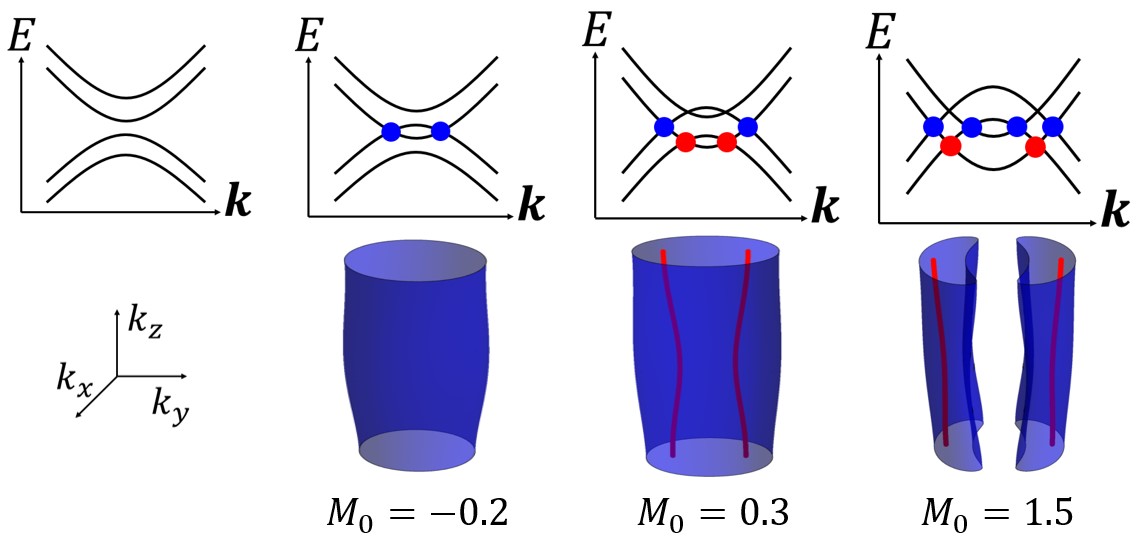}
\caption{Schematic description of double band inversion process and their corresponding nodal structures using the model (\ref{CTmodel}) with $m=1$. Blue surfaces are NSs at $E_F$ and red lines are NLs below $E_F$.}
\label{fig:DBI}
\end{figure}

Even trivial NSs can be transformed to DCNSs via continuous deformation of band structure, which is referred as double band inversion process. Here we provide a simple continuum model for double band inversion, applicable to the systems with inversion and chiral symmetries.

Let us consider a Hamiltonian $H_{\mathrm{conti}}(\mathbf{k})$,
\begin{gather}
H_{\mathrm{conti}}(\mathbf{k})=k_x \sigma_{x}+\left(k_{x}^{2}+k_{y}^{2}-M(k_z)\right)\sigma_{y}\tau_{y}+m\sigma_{x}\tau_{z},
\label{CTmodel}
\end{gather}
where $M(k_z)=M_{0}-0.1\cos k_{z}$. Note that $H_{\mathrm{conti}}(\mathbf{k})$ has symmetries $\mathfrak{T}=\mathcal{K}$ and $\mathcal{C}=\sigma_{z}$. 

While $M_{0}$ increases, we can see that there appears doubly charged nodal surfaces near $k_x=k_y=0$. When $M<-m$, there are nodes neither at $E_F$ nor below $E_F$. After $M_0$ increases so that $M>-m$, a NS appears near $k_x=k_y=0$. When $M$ becomes larger than $0$, there appear two NLs below $E_{F}$ inside the NS. When $M>m^2+1/4$, the NS is separated into two NSs while each NS surrounds one of the NLs, which means that the two NSs are doubly charged. These processes are illustrated in Fig. \ref{fig:DBI}.

\subsection{Lattice model of a class BDI superconductor}

\subsubsection{Constraints on the BdG Hamiltonian}
 In the context of the second quantization, a tight binding Hamiltonian is given by
\begin{gather}
H_{\mathrm{normal}}=\sum_{\alpha\beta\mathbf{k}}\mathcal{H}^{\alpha\beta}_{\mathbf{k}}c^{\alpha\dagger}_{\mathbf{k}}c^{\beta}_{\mathbf{k}},
\end{gather}
where $\alpha$ and $\beta$ are indices for orbital and spin degrees of freedom.
 If $N$ is the number of orbital states used, $\alpha$ and $\beta$ run from 1 to $2N$ respectively. 
 Here $\mathcal{H}_{\mathbf{k}}$ is a $2N\times 2N$ matrix and it describes the band structure.
 Let us call $\mathcal{H}_{\mathbf{k}}$ as a normal state Hamiltonian.
 Note that $\mathcal{H}_{\mathbf{k}}=\mathcal{H}^{\dagger}_{\mathbf{k}}$ due to the hermicity.

We can get a Hamiltonian for a superconductor by adding a pairing field between electrons to the normal state Hamiltonian.
 Under the mean-field approximation, the Hamiltonian to which a pairing potential is added is given by~\cite{altland1997nonstandard}
\begin{gather}
\nonumber H_{\mathrm{SC}}=\sum_{\alpha\beta\mathbf{k}}\left(\mathcal{H}^{\alpha\beta}_{\mathbf{k}}c^{\alpha\dagger}_{\mathbf{k}}c^{\beta}_{\mathbf{k}}+\frac{1}{2}\Delta^{\alpha\beta}_{\mathbf{k}}c^{\alpha\dagger}_{\mathbf{k}}c^{\beta\dagger}_{-\mathbf{k}}\right.
\\\left.+\frac{1}{2}\Delta^{\alpha\beta*}_{-\mathbf{k}}c^{\beta}_{-\mathbf{k}}c^{\alpha}_{\mathbf{k}}\right).
\end{gather}
where $\Delta_{\mathbf{k}}$ is called by a gap function, which is a $2N\times 2N$ matrix.
 The gap function satisfies $\Delta_{\mathbf{k}}=-\Delta^{T}_{-\mathbf{k}}$ due to the fermionic statistics. 
 This Hamiltonian $H_{\mathrm{SC}}$ can be expressed by the Nambu spinor, $\begin{pmatrix}
c^{\alpha}_{\mathbf{k}}\\c^{\alpha\dagger}_{-\mathbf{k}}
\end{pmatrix}$.
 The upper component $c^{\alpha}_{\mathbf{k}}$ of the Nambu spinor is the annihilation operator of an electron with the orbital and spin degree of freedom $\alpha$ and the lower component $c^{\alpha\dagger}_{-\mathbf{k}}$ of it is the annihilation operator of a hole with the orbital and spin degree of freedom $\alpha$.
 Let us call the freedom to choose the electron or hole as the particle-hole degree of freedom.
 
 Then the Hamiltonian $H_{\mathrm{SC}}$ can be rewritten by
\begin{gather}
H_{\mathrm{SC}}=\frac{1}{2}\sum_{\alpha\beta\mathbf{k}}\begin{pmatrix}
c^{\alpha\dagger}_{\mathbf{k}}&&c^{\alpha}_{-\mathbf{k}}
\end{pmatrix}\mathcal{H}_{\mathrm{BdG}}^{\alpha\beta}(\mathbf{k})\begin{pmatrix}
c^{\beta}_{\mathbf{k}}\\c^{\beta\dagger}_{-\mathbf{k}}
\end{pmatrix}.
\end{gather}
Here $\mathcal{H}_{\mathrm{BdG}}(\mathbf{k})$ is a $4N\times 4N$ Bogoliubov–de Gennes(BdG) Hamiltonian, which is given by
\begin{gather}
\mathcal{H}_{\mathrm{BdG}}(\mathbf{k})=\begin{pmatrix}
\mathcal{H}_{\mathbf{k}}&&\Delta_{\mathbf{k}}\\
-\Delta^{*}_{-\mathbf{k}}&&-\mathcal{H}^{T}_{-\mathbf{k}}
\end{pmatrix}.\label{BdGorigin}
\end{gather}
 Due to the particular form of the BdG Hamiltonian, it has a particle-hole symmetry $\mathcal{P}$,
\begin{gather}
\mathcal{P}=\sigma_{x}\mathcal{K},\label{PHsym}
\end{gather}
where $\sigma_{x}$ is a Pauli matrix acting on the particle-hole space.

 If the system has the full spin-rotation symmetry, which is true for class BDI, we can reduce the spin degrees of freedom in the BdG Hamiltonian.
 In the particle-hole space, the spin-rotation symmetry $J_{i}\ (i=x,y,z)$ is given by~\cite{altland1997nonstandard}
\begin{gather}
J_{i}=\begin{pmatrix}
s_{i}&&0\\ 0&&-s_{i}^{T}
\end{pmatrix}.
\end{gather}
Here $s_{i}$ are Pauli matrices and they act on the spin degrees of freedom.
 If the superconducting system has a full spin-rotation symmetry, then $\left[\mathcal{H}_{\mathrm{BdG}},J_{i}\right]=0$ for all $i=x,y,z$.
 This condition changes a form of the BdG Hamiltonian to
\begin{gather}
\mathcal{H}_{\mathrm{BdG}}(\mathbf{k})=\begin{pmatrix}
h_{\mathbf{k}}&&0&&0&&\delta_{\mathbf{k}}\\
0&&h_{\mathbf{k}}&&-\delta_{\mathbf{k}}&&0\\
0&&-\delta^{*}_{-\mathbf{k}}&&-h^{T}_{-\mathbf{k}}&&0\\
\delta^{*}_{-\mathbf{k}}&&0&&0&&-h^{T}_{-\mathbf{k}}
\end{pmatrix},
\end{gather}
where $h_{\mathbf{k}}$ and $\delta_{\mathbf{k}}$ are $N\times N$ matrices and act on the orbital degrees of freedom.
 Switching the second and fourth lows and columns, we can get a block diagonalized form of the BdG Hamiltonian.
 Each block matrix gives the same second quantized Hamiltonian due to the full spin-rotation symmetry.
 The first block is given by
\begin{gather}
\mathcal{H}_{\mathrm{rBdG}}(\mathbf{k})=\begin{pmatrix}
h_{\mathbf{k}}&&\delta_{\mathbf{k}}\\
\delta^{*}_{-\mathbf{k}}&&-h^{T}_{-\mathbf{k}}
\end{pmatrix}.\label{rBdG}
\end{gather}
Let us call this $2N\times 2N$ matrix as a reduced BdG Hamiltonian.
 $h_{\mathbf{k}}$ and $\delta_{\mathbf{k}}$ inherit the hermicity of $\mathcal{H}_{\mathbf{k}}$ and the property of $\Delta_{\mathbf{k}}$ coming from the fermionic statistics; $h^{\dagger}_{\mathbf{k}}=h_{\mathbf{k}}$ and $\delta_{\mathbf{k}}=\delta^{T}_{-\mathbf{k}}$.

The reduced BdG Hamiltonian has a similar form comparing with the BdG Hamiltonian and we can find a particle-hole symmetry $\mathcal{P}_{r}$ for the reduced BdG Hamiltonian, which is given by
\begin{gather}
\mathcal{P}_{r}=ir_{y}\mathcal{K},\label{Pr}
\end{gather}
where $r_{y}$ is a Pauli matrix acting on the reduced particle-hole space.
 We can check that this particle-hole symmetry satisfies $\mathcal{P}_{r}^2=-1$.

To make a BdG Hamiltonian with full spin-rotation symmetry belong to class BDI, we impose three conditions on the BdG Hamiltonian~\cite{bzduvsek2017doubly}; (i) the system has an inversion symmetry $\mathcal{I}$. (ii) it has a time reversal symmetry. (iii) the parity of the gap function is odd.
 The full rotation symmetry of the BdG Hamiltonian imposes the particle-hole symmetry $\tilde{\mathcal{P}}$.
 Considering $\tilde{\mathcal{P}}$ acting on the block diagonalized BdG Hamiltonian whose the first block is the reduced BdG Hamiltonian, we can choose one of the representation of $\tilde{\mathcal{P}}$.
 In this case, $\tilde{\mathcal{P}}$ is given by
\begin{gather}
\tilde{\mathcal{P}}=\begin{pmatrix}
\mathcal{P}_{r} & 0 \\
0 & \mathcal{P}_{r}
\end{pmatrix}.
\end{gather}
Here $\mathcal{P}_{r}$ is the reduced particle-hole symmetry which is given by Eq.~(\ref{Pr}).
 On the other hands, we can consider $\tilde{\mathcal{P}}$ acting on the spin space and particle-hole space.
 In this case, $\tilde{\mathcal{P}}$ is given by
 \begin{gather}
\tilde{\mathcal{P}}=i \sigma_{x}\otimes s_{y} \mathcal{K},\label{Ptilde}
\end{gather}
where $\sigma_{i},s_{j}$ are the Pauli matrices which are acting on the particle hole space and spin space, respectively.

The inversion operator $\mathcal{I}$ of the BdG Hamiltonian can be expressed using the inversion operator $\mathcal{I}_{n}$ of the normal state Hamiltonian. The expression for $\mathcal{I}$ is given by
\begin{gather}
\mathcal{I}=\begin{pmatrix}
\mathcal{I}_{n}&&0\\0&&-\mathcal{I}_{n}
\end{pmatrix}=\sigma_{z}\otimes \mathcal{I}_{n}.\label{Itot}
\end{gather}
Note that the minus sign of the $\mathcal{I}_{n}$ at the second diagonal block comes from the condition (iii).
 Since $\mathcal{I}_{n}$ operates on the normal state Hamiltonian, $\mathcal{I}_{n}$ acts on both the orbital and spin degrees of freedom. 
 On the spin degrees of freedom, however, $\mathcal{I}_{n}$ is a trivial operator. 
 Therefore, $\mathcal{I}_{n}$ is block-diagonalized in the spin degrees of freedom,
\begin{gather}
\mathcal{I}_{n}=s_{0}\otimes\mathcal{I}_{o}.
\end{gather}
$\mathcal{I}_{o}$ is an inversion operator acting on the orbital degrees of freedom. 
 From Eq.~(\ref{Ptilde}) and Eq.~(\ref{Itot}), $(\tilde{\mathcal{P}}\mathcal{I})^{2}=1$.

From $\mathcal{I}\mathcal{H}_{\mathrm{BdG}}(\mathbf{k})\mathcal{I}^{-1}=\mathcal{H}_{\mathrm{BdG}}(-\mathbf{k})$, we can deduce inversion symmetry constraints on $h_{\mathbf{k}}$ and $\delta_{\mathbf{k}}$;
\begin{gather}
\mathcal{I}_{o}h_{\mathbf{k}}\mathcal{I}_{o}^{-1}=h_{-\mathbf{k}},\label{Invh}\\
\mathcal{I}_{o}\delta_{\mathbf{k}}\mathcal{I}_{o}^{-1}=-\delta_{-\mathbf{k}}.\label{Invdel}
\end{gather}

\subsubsection{Lattice model of a 4-band BdG Hamiltonian}
Here, we explain the 4-band lattice model in the main text.
 We first introduce a 4-band BdG Hamiltonian belonging to class BDI on the AA-stacked honeycomb layers.
 The lattice is made by stacking honeycomb layers with same distance and without any translation along in-plane direction [see Fig. 2~(a),~(b) in the main text]. 
 The primitive Bravais vectors are given by
\begin{gather}
\mathbf{R}_{1,2}=\left(\frac{3}{2},\pm\frac{\sqrt{3}}{2},0\right)a,~\mathbf{R}_{3}=(0,0,c),
\end{gather}
where $a$ is a distance between the nearest neighboring atoms in the plane and $c$ is a distance between the layers.
 And relative position vectors $\mathbf{t}_{1,2,3}$ between the nearest neighboring atoms in the plane are given by
\begin{gather}
\mathbf{t}_{1,2}=\left(\frac{1}{2},\pm\frac{\sqrt{3}}{2},0\right)a,~\mathbf{t}_{3}=(-a,0,0).
\end{gather}
For the convenience to write the equations, we define relative position vectors $\mathbf{T}_{1,2,3}$ between the next nearest neighboring atoms in the plane,
\begin{gather}
\mathbf{T}_{1,2}=\left(\pm\frac{3}{2},\frac{\sqrt{3}}{2},0\right)a,~\mathbf{T}_{3}=(0,-\sqrt{3}a,0).
\end{gather}
We add an s orbital at each atomic position which is represented by black dots in Fig. 2~(a),~(b).

We consider the on-site energy $E_{\mathrm{on}}$ which is the same for all s orbitals and the intra-layer nearest-neighbor hopping with an amplitude $t$ and the inter-layer nearest-neighbor hopping with an amplitude $t_{z}$. They produce a normal state Hamiltonian,
\begin{align}
h_{2}(\mathbf{k}) &= \left(E_{\mathrm{on}}+2 t_{z} \cos(k_{z} c)\right)\mathbb{1}_{2\times 2}\nonumber \\
&+ t\sum_{i=1}^{3}\left(\cos (\mathbf{k}\cdot \mathbf{t}_{i})\tau_{x}+\sin (\mathbf{k}\cdot \mathbf{t}_{i})\tau_{y}\right),\label{hnor2}
\end{align}
where $\tau_{i}$ are Pauli matrices acting on the orbital degree of freedom.
 This system has the $C_{6}$ rotation symmetry, the inversion symmetry and the time-reversal symmetry. 
 In particular, the inversion symmetry is represented by $\mathcal{I}=\tau_{x}$ and the time-reversal symmetry is represented by $\mathcal{T}=\mathcal{K}$. 
 If we consider a $2\times 2$ reduced gap function $\delta_{2}(\mathbf{k})$ which is given by
\begin{gather}
\delta_{2}(\mathbf{k})=\psi_{0}\tau_{z},\label{d2}
\end{gather}
where $\psi_{0}$ is a real-valued $s$-wave order parameter, then $\delta_{2}(\mathbf{k})$ is time-reversal symmetric and satisfies Eq.~(\ref{Invdel}).
 This means that the reduced BdG Hamiltonian constituted by $h_{2}(\mathbf{k})$ and $\delta_{2}(\mathbf{k})$ belongs to the class BDI.
 
 This reduced BdG Hamiltonian has two symmetries,
\begin{gather}
\mathfrak{T}=\tau_{x}\otimes r_{0}\mathcal{K},~\mathfrak{B}=i\tau_{x}\otimes r_{y}\mathcal{K},
\end{gather}
where $\tau_{i},r_{j}$ are Pauli matrices and $\tau_{i}$ act on the orbital degrees of freedom and $r_{i}$ act on the reduced particle-hole space.

\begin{figure}[t]
\centering
\includegraphics[width=8.5cm]{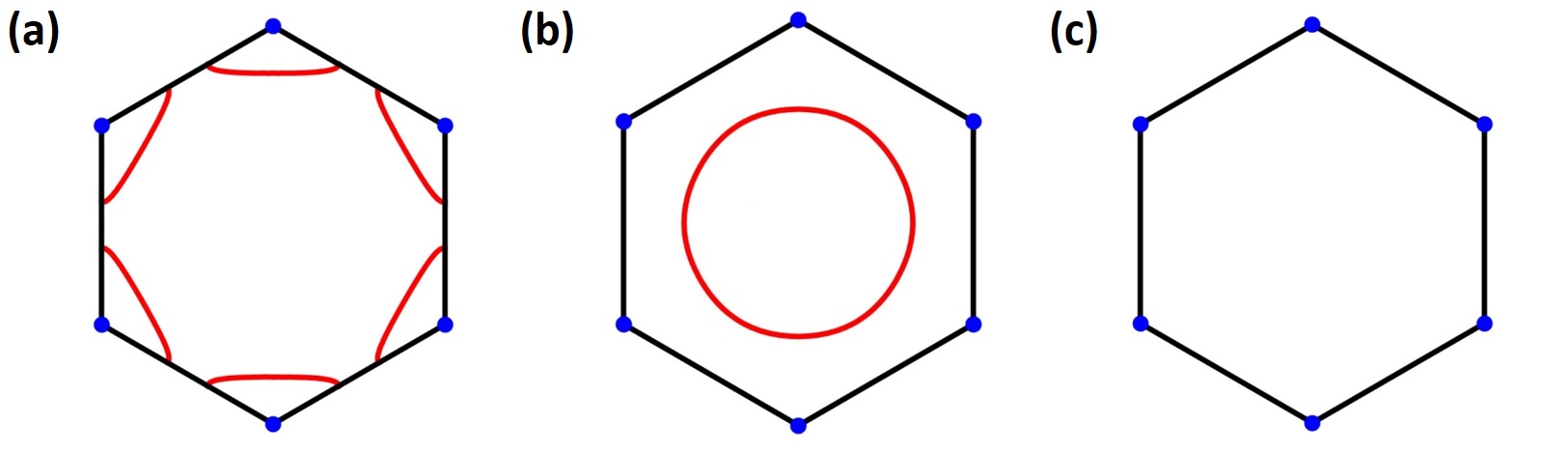}
\caption{Shrinking the nodal surface. The black line is the First Brillouin zone and the blue dot is the nodal line and the red lines are the nodal surface. All of the figures are evaluated on $k_{z}=0$ plane. (a), (b), (c) are evaluated for $E_{0}=-0.6, -1.5, -3$, respectively.}
\label{fig:Shrk}
\end{figure}

In the main text, we consider two sets of the parameters.
 The only difference between the parameters' sets is $E_{\mathrm{on}}$.
 When $E_{\mathrm{on}}$ changes from $-0.5$ to $-3$, while the other parameters are fixed, the nodal surfaces merge and then disappear, see Fig.~\ref{fig:Shrk}.
 We can see that the doubly charged nodal surfaces can be disappeared after merging together, although the doubly charged nodal surface cannot be disappeared alone.
 Note that the nodal surfaces merge at $E_{\mathrm{on}}\approx -0.67$ and the single nodal surface enclosing the BZ center disappears at $E_{\mathrm{on}}\approx -2.8$
 
\subsubsection{Lattice model of a 6-band BdG Hamiltonian}
We expand the $4$ bands BdG Hamiltonian to a $6$ bands BdG Hamiltonian by adding one more s orbital inside the unit cell of the crystal structure of the 4 bands model.
 It is located at the middle of the adjacent honeycomb layers and at the center of the honeycomb structure at the top viewpoint.
 For the added orbital, we consider the on-site energy $E'_{\mathrm{on}}$ and the nearest-neighbor hopping between the added orbital and the orbital in the honeycomb layers with an amplitude $t'$.
 Then $3\times 3$ normal state Hamiltonian $h_{3}(\mathbf{k})$ is given by
\begin{gather}
h_{3}(\mathbf{k})=\left(\begin{array}{ccc}
 & & [h_{3}(\mathbf{k})]_{13} \\
\multicolumn{2}{c}{\smash{\raisebox{.5\normalbaselineskip}{$h_{2}(\mathbf{k})$}}} & [h_{3}(\mathbf{k})]_{23} \\
 \left[h_{3}(\mathbf{k})\right]^{*}_{13} & [h_{3}(\mathbf{k})]^{*}_{23} & [h_{3}(\mathbf{k})]_{33}
\end{array} \right),
\end{gather}
where $h_{2}(\mathbf{k})$ is the $2\times 2$ normal state Hamiltonian before adding the s orbital and the other components of $h_{3}(\mathbf{k})$ are given by
\begin{gather}
[h_{3}(\mathbf{k})]_{13}=t'\sum_{i=1}^{3}(e^{i\mathbf{k}\cdot (\mathbf{t}_{i}+\mathbf{t}_{z})}+e^{i\mathbf{k}\cdot (\mathbf{t}_{i}-\mathbf{t}_{z})}),\\
[h_{3}(\mathbf{k})]_{23}=t'\sum_{i=1}^{3}(e^{-i\mathbf{k}\cdot (\mathbf{t}_{i}+\mathbf{t}_{z})}+e^{-i\mathbf{k}\cdot (\mathbf{t}_{i}-\mathbf{t}_{z})}),\\
[h_{3}(\mathbf{k})]_{33}=E'_{\mathrm{on}},\label{hnor3}
\end{gather}
where $\mathbf{t}_{z}=(0,0,c/2)$. This tight binding Hamiltonian has the inversion symmetry and the time-reversal symmetry. The inversion symmetry operator $\mathcal{I}$ is given by 
\begin{gather}
\mathcal{I}=\begin{pmatrix}
0 & 1 & 0 \\
1 & 0 & 0 \\
0 & 0 & 1
\end{pmatrix},
\end{gather}
and the time-reversal symmetry $\mathcal{T}$ is given by $\mathcal{K}$.
 We consider the time-reversal symmetric reduced gap function $\delta_{3}(\mathbf{k})$ which satisfies Eq.~(\ref{Invdel}),
\begin{gather}
\delta_{3}(\mathbf{k})=\left(\begin{array}{ccc}
 & & \psi'_{0} \\
\multicolumn{2}{c}{\smash{\raisebox{.5\normalbaselineskip}{$\delta_{2}(\mathbf{k})$}}} & -\psi'_{0} \\
 \psi'_{0} & -\psi'_{0} & 0
\end{array} \right),\label{d3}
\end{gather}
where $\delta_{2}(\mathbf{k})$ is the $2\times 2$ gap function given by Eq.~(\ref{d2}) and $\psi'_{0}$ is another $s$-wave order parameter.

\begin{figure}[t]
\centering
\includegraphics[width=8.5cm]{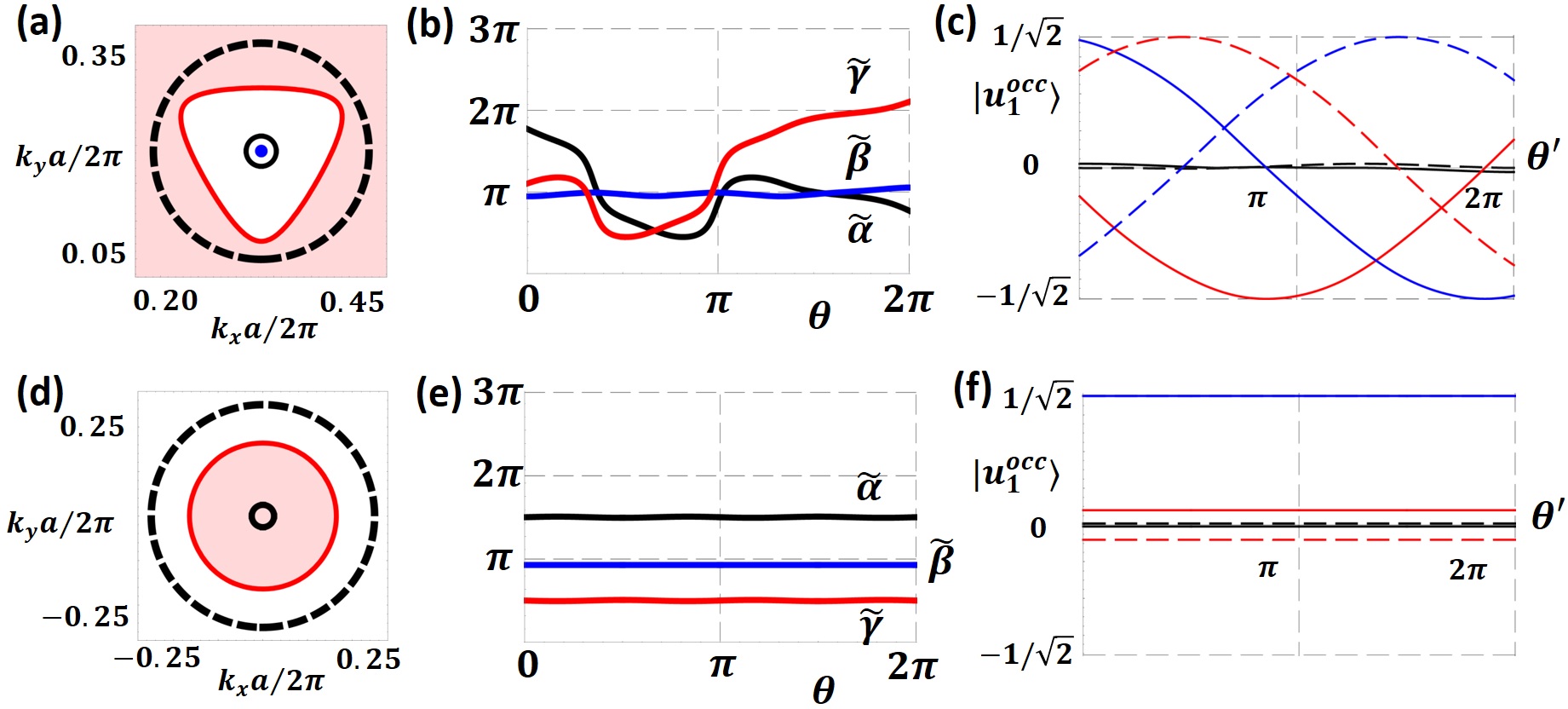}
\caption{NSs and Euler angles for 6-band lattice models.
(a) A cross section of the NS with $c_{\textrm{1D}}=1$ on the $k_{z}=0$ plane. 
The light red (white) region has the same meaning as in Fig.~2. 
(b) Lifted Euler angles of the flattened Hamiltonian along the dashed black lines in (a). 
(c) Components of $|u^{\mathrm{occ}}_{1}\rangle$ along the small black circle in (a). 
Red, blue and black solid (dashed) curves correspond to the first, second and third (fourth, fifth and sixth) components of $|u^{\mathrm{occ}}_{1}\rangle$.
(d-f) Similar figures for the NS with $c_{\textrm{1D}}=0$.}
\label{fig:BDI6B}
\end{figure}

To evaluate $c_{\textrm{1D}}$ in the 6-band model, we compute the homotopy equivalence class of $A_{FB}(\mathbf{k})\in \mathrm{O(3)}$ defined on a circle surrounding a NS. 
$A_{FB}$ can be restricted to $\mathrm{SO(3)}$ using a map $f : \left[\mathrm{O}(3)-\mathrm{SO}(3)\right]\rightarrow \mathrm{SO}(3)$ as before. 
In terms of three Euler angles $\alpha, \beta, \gamma\in [0,2\pi]$, $A_{FB}$ can be writtened as
$A_{FB}= e^{\gamma L_{3}}e^{\beta L_{1}}e^{\alpha L_{3}}$
where $L_{i}~(i=1,2,3)$ are generators of Lie algebra $\mathfrak{so}(3)$ defined by $[L_{i}]_{jk}=-\epsilon_{ijk}$.\\
\indent To determine the homotopy equivalence class of a closed loop in $\mathrm{SO(3)}$, we examine a lifting of the closed loop to the double covering group $\mathrm{SU(2)}$ by replacing $L_{j}$ by $-\frac{i}{2}\sigma_{j}$ where $\sigma_{j}$ are Pauli matrices~\cite{bzduvsek2019nonabelian}, and substituting $\alpha, \beta, \gamma$ with $\tilde{\alpha}, \tilde{\beta}, \tilde{\gamma}\in [0,4\pi]$, respectively. 
The lifted loop can take a form of either a closed loop or an open line because a covering map $p:\mathrm{SU(2)}\rightarrow \mathrm{SO(3)}$ is two-to-one.
In the former case, as $\mathrm{SU(2)}$ is simply connected, a closed loop in $\mathrm{SU(2)}$ can always be contracted to a point. 
This means that the homotopy equivalence class of the original loop is a trivial element of $\pi_{1}(\mathrm{SO(3)})$. 
In the latter case, on the other hand, as the two end points of the lifted open line in $\mathrm{SU(2)}$ have a fixed $L^{2}$-norm $2\sqrt{2}$, it cannot be smoothly contracted to a point
by deforming the original loop in $\mathrm{SO(3)}$ continuously.
This corresponds to the case when the original loop corresponds to a non-trivial element of $\pi_{1}(\mathrm{SO(3)})$.
As $\pi_{1}(\mathrm{SO(3)})=\mathbb{Z}_{2}$, the shape of the lifted line (closed or open) gives sufficient information for the homotopy equivalence class of the original loop. \\ 
\indent Fig.~\ref{fig:BDI6B}(b, e) show the lifted Euler angles of $A_{FB}(\mathbf{k})$ for the 6-band model with the NSs in Fig.~\ref{fig:BDI6B}(a, d).
The case with different (same) lifted Euler angles at $\theta =0$ and $\theta =2\pi$ corresponds to $c_{\textrm{1D}}=1$ ($c_{\textrm{1D}}=0$).
Here $\theta$ parametrizes the black dotted circle. 
In Fig.~\ref{fig:BDI6B}(c, f), we plot the components of the highest occupied states $|u^{\mathrm{occ}}_{1}\rangle$
computed on a small black circle parametrized by $\theta'\in[0,2\pi]$ inside the NS shown in Fig.~\ref{fig:BDI6B}(a, d).
The opposite signs of $|u^{\mathrm{occ}}_{1}\rangle$ at $\theta'=0$ and $2\pi$ in Fig.~\ref{fig:BDI6B}(c) 
indicate the presence of a NL between occupied bands, marked by a blue dot in Fig.~\ref{fig:BDI6B}(a), which again confirms
the linking structure of DCNSs.
 
\subsubsection{1D charge calculation of a 2N-band model}
The 1D charge of the 2N-band case is defined by the homotopy equivalence class of $A_{FB}(\mathbf{k})\in \mathrm{SO(N)}$.
 Similar to $\mathrm{SO(3)}$, an arbitrary element of $\mathrm{SO(N)}$ can be represented by generalized $\mathrm{N(N-1)/2}$ Euler angles.
 Here, We consider the passive transformation.
 $A_{FB}(\mathbf{k})$ can be determined by specifying how $A_{FB}(\mathbf{k})$ transforms the standard orthonormal basis $\{\mathbf{e}_{i}\}$.
 The transformed basis $\{\mathbf{e}^{\mathrm{N}}_{j}\}$ is given by
\begin{gather}
\mathbf{e}^{\mathrm{N}}_{j}=\sum_{i=1}^{\mathrm{N}}\mathbf{e}_{i}[A_{FB}(\mathbf{k})]_{ij}.
\end{gather}
\indent We are going to represent $A_{FB}(\mathbf{k})$ by a product of rotation matrices $\mathrm{M}_{i}\in\mathrm{SO(N)}$ $(i=1,\cdots,\mathrm{N-1})$.
 The first rotation matrix $\mathrm{M}_{1}$ turns the basis $\{\mathbf{e}_{i}\}$ into the other basis $\{\mathbf{e}^{1}_{i}\}$.
 And we are going to set $\mathrm{M}_{1}$ to make the last rotated basis vector $\{\mathbf{e}^{1}_{\mathrm{N}}\}$ satisfy
\begin{gather}
\mathbf{e}^{1}_{\mathrm{N}}=\mathbf{e}^{\mathrm{N}}_{\mathrm{N}}.
\end{gather}
The second rotation matrix $\mathrm{M}_{2}$ turns the basis $\{\mathbf{e}^{1}_{i}\}$ into the other basis $\{\mathbf{e}^{2}_{i}\}$.
 And we are going to set $\mathrm{M}_{2}$ to make the rotated basis vectors $\{\mathbf{e}^{2}_{\mathrm{N}}\}$ and $\{\mathbf{e}^{2}_{\mathrm{N-1}}\}$ satisfy 
\begin{gather}
\mathbf{e}^{2}_{\mathrm{N}}=\mathbf{e}^{\mathrm{N}}_{\mathrm{N}},\mathbf{e}^{2}_{\mathrm{N-1}}=\mathbf{e}^{\mathrm{N}}_{\mathrm{N-1}}.
\end{gather}
Like this, we will define all the matrices $\mathrm{M}_{i}$ so that
\begin{gather}
A_{FB}(\mathbf{k})=\mathrm{M}_{\mathrm{N-1}}\cdots \mathrm{M}_{2}\mathrm{M}_{1}.\label{Mprod}
\end{gather}
Let us consider $\mathrm{M}_{1}$ first. 
 The N-dimensional unit vector $\mathbf{e}^{\mathrm{N}}_{\mathrm{N}}$ can always be defined by $\mathrm{N}-1$ angles $\alpha^{\mathrm{N}}_{1},\cdots,\alpha^{\mathrm{N}}_{\mathrm{N-1}}$,
\begin{align}
\mathbf{e}^{\mathrm{N}}_{\mathrm{N}}&=\sin\alpha^{\mathrm{N}}_{1}\sin\alpha^{\mathrm{N}}_{2}\cdots\sin\alpha^{\mathrm{N}}_{\mathrm{N}-1}\mathbf{e}_{1}\nonumber \\
&+\cos\alpha^{\mathrm{N}}_{1}\sin\alpha^{\mathrm{N}}_{2}\cdots\sin\alpha^{\mathrm{N}}_{\mathrm{N}-1}\mathbf{e}_{2}\nonumber \\
&+\cos\alpha^{\mathrm{N}}_{2}\cdots\sin\alpha^{\mathrm{N}}_{\mathrm{N}-1}\mathbf{e}_{3}\nonumber \\
&+\cdots+\cos\alpha^{\mathrm{N}}_{\mathrm{N}-1}\mathbf{e}_{\mathrm{N}}.\label{Nangexp}
\end{align}
Let us define the $\mathrm{N(N-1)/2}$ generators $L_{ij}$ of the Lie algebra $\mathfrak{so}\mathrm{(N)}$ by
\begin{gather}
[L_{ij}]_{nm}=
\begin{cases}
-1, & \mathrm{if}~i=n,~j=m, \\
1, & \mathrm{if}~i=m,~j=n, \\
0, & \mathrm{otherwise},
\end{cases}
\end{gather}
where $i<j$.
 Note that the matrix exponential $e^{\theta L_{ij}}$ of a generator $L_{ij}$ is an element of the Lie group $\mathrm{SO(N)}$ and $e^{\theta L_{ij}}$ is a counter-clockwise rotation of the basis vectors $\mathbf{e}_{i},\mathbf{e}_{j}$ through angle $\theta$.
 If we define $\mathrm{M}_{1}$ by
\begin{gather}
\mathrm{M}_{1}=e^{-\alpha^{\mathrm{N}}_{1}L_{12}}e^{-\alpha^{\mathrm{N}}_{2}L_{23}}\cdots e^{-\alpha^{\mathrm{N}}_{\mathrm{N-1}}L_{\mathrm{N}-1,\mathrm{N}}},
\end{gather}
one can show that 
\begin{gather}
\mathbf{e}^{1}_{\mathrm{N}}=\sum_{i=1}^{\mathrm{N}}\mathbf{e}_{i}[\mathrm{M}_{1}]_{i\mathrm{N}}=\mathbf{e}^{\mathrm{N}}_{\mathrm{N}}.
\end{gather}
\indent To define $\mathrm{M}_{2}$, we are going to represent $\mathbf{e}^{\mathrm{N}}_{\mathrm{N-1}}$ using the basis $\{\mathbf{e}_{i}^{1}\}$ first. 
 Since $\mathbf{e}^{\mathrm{N}}_{\mathrm{N-1}}$ and $\mathbf{e}^{\mathrm{N}}_{\mathrm{N}}$ are orthogonal, $\mathbf{e}^{\mathrm{N}}_{\mathrm{N-1}}$ can be written by the linear combination of the $\mathrm{N-1}$ basis vectors $\mathbf{e}^{1}_{1},\cdots,\mathbf{e}^{1}_{\mathrm{N-1}}$.
 Similar to Eq.~(\ref{Nangexp}), $\mathbf{e}^{\mathrm{N}}_{\mathrm{N-1}}$ can be defined by $\mathrm{N-2}$ angles $\alpha^{\mathrm{N-1}}_{1},\cdots,\alpha^{\mathrm{N-1}}_{\mathrm{N-2}}$,
\begin{align}
\mathbf{e}^{\mathrm{N}}_{\mathrm{N-1}}=&\sin\alpha^{\mathrm{N-1}}_{1}\sin\alpha^{\mathrm{N-1}}_{2}\cdots\sin\alpha^{\mathrm{N-1}}_{\mathrm{N-2}}\mathbf{e}^{1}_{1}\nonumber \\
+&\cos\alpha^{\mathrm{N-1}}_{1}\sin\alpha^{\mathrm{N-1}}_{2}\cdots\sin\alpha^{\mathrm{N-1}}_{\mathrm{N-2}}\mathbf{e}^{1}_{2}\nonumber \\
+&\cos\alpha^{\mathrm{N-1}}_{2}\cdots\sin\alpha^{\mathrm{N-1}}_{\mathrm{N-2}}\mathbf{e}^{1}_{3}\nonumber \\
+&\cdots+\cos\alpha^{\mathrm{N-1}}_{\mathrm{N-2}}\mathbf{e}^{1}_{\mathrm{N-1}}.
\end{align}
$\mathrm{M}_{2}$ can be defined by a similar form of $\mathrm{M}_{1}$,
\begin{gather}
\mathrm{M}_{2}=e^{-\alpha^{\mathrm{N-1}}_{1}L_{12}}e^{-\alpha^{\mathrm{N-1}}_{2}L_{23}}\cdots e^{-\alpha^{\mathrm{N-1}}_{\mathrm{N-2}}L_{\mathrm{N-2},\mathrm{N-1}}},\label{M2}
\end{gather}
And one can show that
\begin{gather}
\mathbf{e}^{2}_{\mathrm{N-1}}=\sum_{i=1}^{\mathrm{N}}\mathbf{e}^{1}_{i}[\mathrm{M}_{2}]_{i,\mathrm{N-1}}=\mathbf{e}^{\mathrm{N}}_{\mathrm{N-1}}.
\end{gather}
Moreover, $\mathrm{M_{2}}$ in Eq.~(\ref{M2}) does not rotate $\mathrm{N}$-th basis vector.
 This means that
\begin{gather}
\mathbf{e}^{2}_{\mathrm{N}}=\mathbf{e}^{1}_{\mathrm{N}}=\mathbf{e}^{\mathrm{N}}_{\mathrm{N}}.
\end{gather}
\indent Following these steps, we get the rotation matrices $\mathrm{M}_{1},\cdots,\mathrm{M}_{\mathrm{N-1}}$ satisfying Eq.~(\ref{Mprod}).
 The $\mathrm{N(N-1)/2}$ angles $\alpha^{i}_{j}~(i>j)$ are the generalized Euler angles of $\mathrm{SO(N)}$~\cite{hoffman1972EulerangSO(N)}.\\
\indent The next step is lifting $A_{FB}(\mathbf{k})$ in $\mathrm{SO(N)}$ to $\mathrm{Spin(N)}$ which is a doubly covering space of $\mathrm{SO(N)}$.
 Lifting $A_{FB}(\mathbf{k})$ replaces $\alpha^{i}_{j}$ in each $\mathrm{M}_{k}$ by $\tilde{\alpha}^{i}_{j}\in[0,4\pi]$. 
 Also, we have to change $L_{ij}$ in each $\mathrm{M}_{k}$ into the generators $t_{ij}$ of the Lie algebra $\mathfrak{spin}\mathrm{(N)}$,
\begin{gather}
t_{ij}=-\frac{1}{4}[\Gamma_{i},\Gamma_{j}],
\end{gather}
where $\Gamma_{i}~(i=1,\cdots,\mathrm{N})$ are gamma matrices of dimensions $2^{[\mathrm{N}/2]}\times 2^{[\mathrm{N}/2]}$ which mutually anti-commute, i.e. $\{\Gamma_{i},\Gamma_{j}\}=2\delta_{ij}$~\cite{bzduvsek2019nonabelian}.

\section{Class D}
In class D, we have $\mathfrak{B}^2=1$.
 We represent this symmetry as a complex conjugation operator $\mathcal{K}$.
 From this symmetry, we can deduce the nodal structure of class D.
 Consider an effective 2-band Hamiltonian,
\begin{align}
H_{\mathrm{eff}}(\mathbf{k})=E_{0}(\mathbf{k})\mathbb{1}_{2\times 2}+h_{x}(\mathbf{k})\sigma_{x}+h_{y}(\mathbf{k})\sigma_{y}+h_{z}(\mathbf{k})\sigma_{z},
\end{align}
which describing the neighborhood of a node.
 Here $E_{0}(\mathbf{k}),h_{x,y,z}(\mathbf{k})$ are real-valued functions.
 Like class BDI, there is a symmetry which is anti-commuting with a Hamiltonian: $\{H(\mathbf{k}),\mathfrak{B}\}=0$.
 This makes a nodal structure at $E_{0}=0$ and that at $E_{0}\neq 0$ different. 
 
If $E_{0}=0$, $h_{x}=h_{z}=0$ due to $\mathfrak{B}=\mathcal{K}$.
 Then there is one constraint on the node, which is $h_{y}=0$.
 Therefore, the node at the Fermi level is a surface in 3D momentum space.
 
If $E_{0}\neq 0$, this type of nodes should appear between occupied bands or unoccupied bands. 
 Since $\mathfrak{B}$ is anti-commute with the Hamiltonian, $\mathfrak{B}$ does not constrain on the nodal structure at $E_{0}\neq 0$.
 This means that a node between occupied bands or unoccupied bands is a point.
\\
\subsection{Topological charges of class D}
Focusing on the node at $E_{F}$, we can find two types of topological charges; a 0D charge and a 2D charge.
 The 0D charge is defined by the Pfaffian of the Hamiltonian. 
 Due to the symmetry $\mathfrak{B}$, the Hamiltonian $H$ is purely imaginary. 
 Then $H$ should be skew-symmetric,
\begin{gather}
H^{T}=H^{*}=-H.
\end{gather}
Since $\mathfrak{B}$ makes the Hamiltonian even dimensional, $iH$ is a skew-symmetric and even-dimensional real matrix.

For a skew-symmetric and even-dimensional matrix, we can define the Pfaffian of the matrix~\cite{bzduvsek2017doubly}. 
For a given $2N\times 2N$ skew-symmetric matrix $A$, we can always find an orthogonal matrix $Q$ such that $A=Q^{T}\Sigma Q$ where
\begin{gather}
\Sigma=\bigoplus_{n=1}^{N}\begin{pmatrix}
0&&a_{n}\\-a_{n}&&0
\end{pmatrix}.
\end{gather}
In that case, the Pfaffian $\mathrm{Pf}(A)$ of $A$ is given by $\prod_{n=1}^{N}a_{n}$. Since the flat-band Hamiltonian $H_{\mathrm{FB}}$ is also a skew-symmetric and even-dimensional matrix, $\mathrm{Pf}\left[iH_{\mathrm{FB}}\right]$ can be defined.

The Pfaffian $\mathrm{Pf}[iH_{\mathrm{FB}}]$ is related to the determinant $\mathrm{det}[iH_{\mathrm{FB}}]$ of the flat-band Hamiltonian through an identity,
\begin{gather}
\mathrm{det}[iH_{\mathrm{FB}}]=\left(\mathrm{Pf}[iH_{\mathrm{FB}}]\right)^2.
\end{gather}

\noindent For the $H_{\mathrm{FB}}$ with $2N$ bands, $\mathrm{det}[H_{\mathrm{FB}}]=(-1)^{N}$ because there are $N$ bands with an energy $1$ and the other $N$ bands with an energy $-1$. Therefore, $\mathrm{Pf}[iH_{\mathrm{FB}}]$ should be $\pm 1$. And the boundary of each sector with $\mathrm{Pf}[iH_{\mathrm{FB}}]=\pm 1$ appears when the gap is closed. From these facts, we can define the 0D charge of the node by
\begin{align}
c_{\mathrm{D}}(S^{0})=\mathrm{Pf}\left[iH_{\mathrm{FB}}(\mathbf{k}_{1})\right]\cdot\mathrm{Pf}\left[iH_{\mathrm{FB}}(\mathbf{k}_{2})\right]\in \{-1,1\},
\end{align}
where $S^{0}=\{\mathbf{k}_{1},\mathbf{k}_{2}\}$ and $\mathbf{k}_{1}$ and $\mathbf{k}_{2}$ locate on the opposite sides of the node. 

Similar to class BDI, the 0D charge can be explained by a band inversion across the nodal surface.
 For an occupied state $|u^{\mathrm{occ}}_{n\mathbf{k}}\rangle$ with an energy $-E_{n\mathbf{k}}$, there is an unoccupied state $|u^{\mathrm{unocc}}_{n\mathbf{k}}\rangle\propto \mathfrak{B}|u^{\mathrm{occ}}_{n\mathbf{k}}\rangle$ with an energy $E_{n\mathbf{k}}$.
 We can express $|u_{n}^{\mathrm{occ}}\rangle$ and $|u_{n}^{\mathrm{unocc}}\rangle$ by
\begin{gather}
|u_{n}^{\mathrm{occ}}\rangle =\frac{1}{\sqrt{2}}\left(|u_{n}^{\mathrm{r}}\rangle+i|u_{n}^{\mathrm{i}}\rangle\right),\label{Docc}\\
|u_{n}^{\mathrm{unocc}}\rangle =\frac{1}{\sqrt{2}}\left(|u_{n}^{\mathrm{r}}\rangle-i|u_{n}^{\mathrm{i}}\rangle\right)\label{Dunocc},
\end{gather}
where $|u_{n}^{\mathrm{r}}\rangle$ and $|u_{n}^{\mathrm{i}}\rangle$ are real-valued vectors.
 It follows from the orthonormal condition of the eigenstates that $\left\{|u_{n}^{\mathrm{r}}\rangle,|u_{n}^{\mathrm{i}}\rangle:n=1,\cdots N\right\}$ satisfies the orthonormal condition, too.
 Changing the basis from $\{|u^{\mathrm{occ}}_{n}\rangle,|u^{\mathrm{unocc}}_{n}\rangle\}$ to $\left\{|u_{n}^{\mathrm{r}}\rangle,|u_{n}^{\mathrm{i}}\rangle\right\}$, we can get
\begin{gather}
iH_{\mathrm{FB}}=\sum_{n=1}^{N} \left(|u_{n}^{\mathrm{i}}\rangle\langle u_{n}^{\mathrm{r}}|-|u_{n}^{\mathrm{r}}\rangle\langle u_{n}^{\mathrm{i}}|\right).
\end{gather}
In this basis, $iH_{\mathrm{FB}}$ is block-diagonalized.
 If we set an order of basis by $\{|u_{n}^{\mathrm{r}}\rangle,|u_{n}^{\mathrm{i}}\rangle\}$ in each block, $iH_{\mathrm{FB}}$ is given by $N$-direct sum of $\begin{pmatrix}
0&&1\\-1&&0
\end{pmatrix}$.
 Therefore, $\mathrm{Pf}\left[iH_{\mathrm{FB}}\right]=1$.

Suppose that there is a band inversion between the highest occupied band and lowest unoccupied band across the nodal surface. This results in adding a minus sign to $|u_{1}^{\mathrm{i}}\rangle$. Then $iH'_{\mathrm{FB}}$ after a band inversion is given by
\begin{gather}
iH'_{\mathrm{FB}}=O^{T}\cdot iH_{\mathrm{FB}}\cdot O,\label{iHbasis}
\end{gather}
where $O$ is an orthogonal matrix which changes the sign of $|u_{1}^{i}\rangle$.
 The well-known identity of Pfaffian is given by
\begin{gather}
\mathrm{Pf}[O^{T}\cdot iH_{\mathrm{FB}}\cdot O]=\mathrm{det}O\cdot \mathrm{Pf}[iH_{\mathrm{FB}}],\label{Pforder}
\end{gather}
which shows that $\mathrm{Pf}[iH_{\mathrm{FB}}]$ has to change its sign across the nodal surface when there is a band inversion.
 Note that, across the nodal surface, the only allowed change for eigenstates is a band inversion between the highest occupied state and lowest unoccupied state. The main reason is almost the same as that of class BDI: the symmetry $\mathfrak{B}$ does not allow to mix the highest occupied state and lowest unoccupied state across the nodal surface.
 
The 2D charge of class D is defined by the Chern number.
 Consider a sphere $S^{2}$ surrounding a nodal surface.
 Then the 2D charge $c_{\mathrm{D}}(S^{2})$ of the nodal surface is given by
\begin{gather}
c_{\mathrm{D}}(S^2)=\frac{i}{2\pi} \sum_{n\in \mathrm{occ}}\oint_{S^2} d^{2}\mathbf{k} \cdot \nabla_{\mathbf{k}} \times \mathbf{A}^{\mathrm{occ}}_{n}(\mathbf{k}),
\end{gather}
where $\mathbf{A}^{\mathrm{occ}}_{n}(\mathbf{k})=\langle u^{\mathrm{occ}}_{n}(\mathbf{k})|\nabla_{\mathbf{k}}|u^{\mathrm{occ}}_{n}(\mathbf{k})\rangle$ is the Berry connection of $n$-th occupied band~\cite{bzduvsek2017doubly}.
 We can express the 2D charge as a sum of the Chern numbers of each occupied band,
\begin{gather}
c_{\mathrm{D}}(S^2)=\sum_{n\in \mathrm{occ}} c^{\mathrm{occ}}_{\mathrm{A},n}(S^2),\label{2Din}
\end{gather}
where $c^{\mathrm{occ}}_{\mathrm{A},n}(S^2)$ is the Chern number of the $n$-th occupied band over the sphere $S^{2}$. 

The topological charge of a nodal point between the occupied bands can be defined using Chern numbers of each band.
 For a nodal point between $n$-th and $n+1$-th occupied bands, we can get non-trivial Chern numbers  $c^{\mathrm{occ}}_{\mathrm{A},n}(S^{2})$ and $c^{\mathrm{occ}}_{\mathrm{A},n+1}(S^{2})$ of $n$-th and $n+1$-th occupied bands, where $S^{2}$ surrounds the nodal point.

\subsection{Linking structure of class D}
The 2D charge of a nodal surface is given by
\begin{gather}
c_{D}(S^{2}_{\mathrm{out}})=\sum_{n\in \mathrm{occ}} c^{\mathrm{occ}}_{A,n}(S^{2}_{\mathrm{out}}),
\end{gather}
where $S^{2}_{\mathrm{out}}$ is a sphere surrounding the nodal surface. For $n\geq 2$, the $n$-th occupied band is continuous across the nodal surface. This means that
\begin{gather}
c^{\mathrm{occ}}_{\mathrm{A},n}(S^{2}_{\mathrm{out}})=c^{\mathrm{occ}}_{\mathrm{A},n}(S^{2}_{\mathrm{in}}),
\end{gather}
where $S^{2}_{\mathrm{in}}$ is a sphere inside the nodal surface. However, the case of $n=1$ is different. Due to the band inversion across the nodal surface, the Chern number of the highest occupied band is switched with that of the lowest unoccupied band,
\begin{gather}
c^{\mathrm{occ}}_{\mathrm{A},1}(S^{2}_{\mathrm{out}})=c^{\mathrm{unocc}}_{\mathrm{A},1}(S^{2}_{\mathrm{in}}).
\end{gather}
Using the symmetry $\mathfrak{B}$, we can deduce a simple relation between the Chern number of the unoccupied band and that of the occupied band. Since $|u^{\mathrm{unocc}}_{n\mathbf{k}}\rangle\propto \mathfrak{B}|u^{\mathrm{occ}}_{n\mathbf{k}}\rangle$, the Berry connection of the $n$-th occupied band is related to that of the $n$-th unoccupied band by
\begin{gather}
\left(\mathbf{A}^{\mathrm{occ}}_{n}\right)^{*}=\mathbf{A}^{\mathrm{unocc}}_{n}.\label{Aoccunocc}
\end{gather}
Also we can deduce that $\mathbf{A}^{\mathrm{occ}}_{n}+\left(\mathbf{A}^{\mathrm{occ}}_{n}\right)^{*}=\nabla_{\mathbf{k}}\langle u^{\mathrm{occ}}_{n}|u^{\mathrm{occ}}_{n}\rangle=0$ from the normalization of the occupied state. The above two equations say that the Chern number of the $n$-th occupied band has a different sign comparing with that of the $n$-th unoccupied band,
\begin{gather}
c_{\mathrm{A},n}^{\mathrm{unocc}}(S^{2})=-c_{\mathrm{A},n}^{\mathrm{occ}}(S^{2}).\label{cAoccunocc}
\end{gather}
\indent To sum up, the 2D charge of the nodal surface is given by
\begin{gather}
c_{\mathrm{D}}(S^{2}_{\mathrm{out}})=
-c_{\mathrm{A},1}^{\mathrm{occ}}(S^{2}_{\mathrm{in}})+\sum_{n\in \mathrm{occ}-\{1\}} c^{\mathrm{occ}}_{\mathrm{A},n}(S^{2}_{\mathrm{in}}).
\end{gather}
Inside the nodal surface, $\sum_{n\in \mathrm{occ}} c^{\mathrm{occ}}_{\mathrm{A},n}(S^{2}_{\mathrm{in}})=0$ because there is no nodal surface inside $S^{2}_{\mathrm{in}}$. Consequently,
\begin{gather}
c_{\mathrm{D}}(S^{2}_{\mathrm{out}})=-2 c^{\mathrm{occ}}_{\mathrm{A},1}(S^{2}_{\mathrm{in}}).\label{Dlinkingeq}
\end{gather}
Therefore, the 2D charge of the nodal surface comes from the Chern number of the highest occupied band inside the node~\cite{bzduvsek2017doubly}.

\begin{figure}[t]
\includegraphics[width=\linewidth]{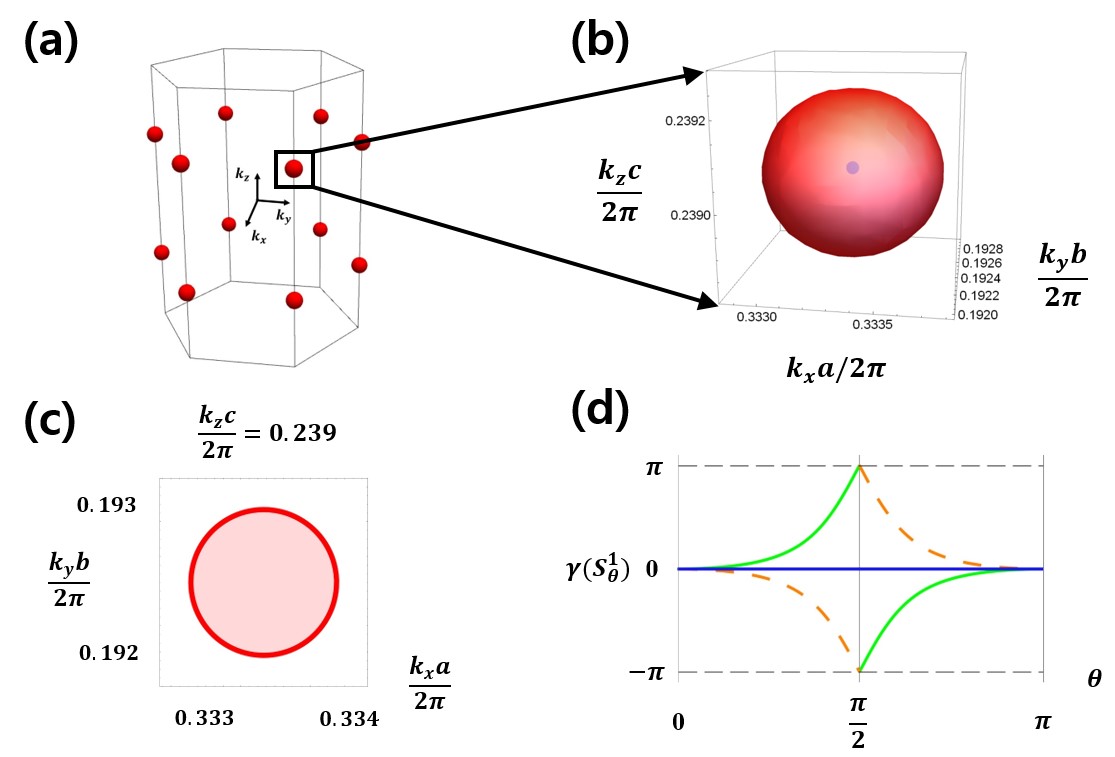}
\caption{Topological charges of the nodes of the 8 bands BdG Hamiltonian belonging to class D. (a) The first Brillouin zone and the nodal surface. The red surfaces are the nodal surface. (b) One of the nodal surface. The blue dot inside the surface is a nodal point between the topmost occupied band and the second topmost occupied band. (c) A cross section of nodal structure of (a) at $k_{z}c/2\pi=0.293$. The light red(white) region is where the Pfaffian of $iH_{\mathrm{FB}}$ is positive(negative). (d) Winding of the Berry phase on a sphere surrounding the nodal surface in (b). $\gamma(S^{1}_{\theta})$ is the Berry phase of a band along a parallel of latitude with a latitude $\theta$. The values of straight lines are calculated outside the nodal surface and those of the dashed line are calculated inside the nodal surface. The green line corresponds to the topmost occupied band and the second topmost occupied band and the blue line corresponds to the third topmost occupied band and the fourth topmost occupied band. The dashed orange line corresponds to the topmost occupied band.}
\label{fig:Dcharge}
\end{figure}

\subsection{Lattice model of a class D superconductor}
We start from a BdG Hamiltonmian (\ref{BdGorigin}). To make a BdG Hamiltonian belonging to class D, we impose four conditions on a BdG Hamiltonian; (i) the system has an inversion symmetry. (ii) a parity of a gap function is even. (iii) there is no time-reversal symmetry. (iv) there is no spin-rotation symmetry~\cite{bzduvsek2017doubly}. From the conditions (i) and (ii), we can deduce that
\begin{gather}
\mathcal{I}\mathcal{H}_{\mathbf{k}}\mathcal{I}^{-1}=\mathcal{H}_{-\mathbf{k}},\\
\mathcal{I}\Delta_{\mathbf{k}}\mathcal{I}^{-1}=\Delta_{-\mathbf{k}},
\end{gather}
where $\mathcal{I}$ is an inversion symmetry operator of the normal state Hamiltonian.
 From the condition (i), the BdG Hamiltonian has a symmetry $\mathfrak{B}=\mathcal{PI}$, where $\mathcal{P}$ is the particle-hole symmetry which is Eq. (\ref{PHsym}).
 From the conditions (ii) and (iii), $\mathfrak{B}^{2}=1$.

We made a simple tight binding model to check the linking structure in class D. The tight binding model is constructed on the AA-stacked honeycomb layers which is the same lattice as 4 bands modal of class BDI.
 An s orbital is located at each atomic position and all the orbitals are the same.
 Inside the layer, we consider the on-site energy, the nearest-neighbor hopping and the next nearest-neighbor hopping with the amplitudes $E_{\mathrm{on}}$, $t_{1}$, $t_{2}$, respectively.
 These parameters make a tight binding Hamiltonian $\mathcal{H}_{1}(\mathbf{k})$,
\begin{align}
\mathcal{H}_{1}(\mathbf{k})=&(E_{\mathrm{on}}+2t_{2}\sum_{i=1}^{3} \left(\cos (\mathbf{k}\cdot \mathbf{T}_{i})\right) \sigma_{0}\otimes\tau_{0}\nonumber\\
+& t_{1} \sigma_{0}\otimes \sum_{i=1}^{3}\left(\cos (\mathbf{k}\cdot \mathbf{t}_{i})\tau_{x}+\sin (\mathbf{k}\cdot \mathbf{t}_{i})\tau_{y}\right),
\end{align}
where $\sigma_{i},\tau_{j}$ are the Pauli matrices and act on the spin degree of freedom and the orbital degree of freedom, respectively.
 We also consider the spin-orbit coupling between the next nearest neighboring atoms with an amplitude $v_{z}$.
 This makes another tight binding Hamiltonian $\mathcal{H}_{2}(\mathbf{k})$,
\begin{gather}
\mathcal{H}_{2}(\mathbf{k})=v_{z}\sum_{i=1}^{3} \sin (\mathbf{k}\cdot\mathbf{T}_{i}) \sigma_{z}\otimes\tau_{z}.
\end{gather}
Due to this term, the only left spin-rotation symmetry is $\sigma_{z}$.
 Between the layers, we only consider the nearest-neighbor hopping with an amplitude $t_{z}$ and the tight binding Hamiltonian $\mathcal{H}_{3}(\mathbf{k})$ resulting from $t_{z}$ is given by
\begin{gather}
\mathcal{H}_{3}(\mathbf{k})=2 t_{z} \cos(\mathbf{k}\cdot\mathbf{R}_{3}) \sigma_{0}\otimes\tau_{0}.
\end{gather}

The total tight binding Hamiltonian $\mathcal{H}(\mathbf{k})=\sum_{i=1}^{3} \mathcal{H}_{i}(\mathbf{k})$ has the inversion symmetry $\mathcal{I}$, the time-reversal symmetry $\mathcal{T}$ and $\mathrm{U}(1)$ spin-rotation symmetry $\sigma_{z}$. The inversion symmetry and time-reversal symmetry are represented by
\begin{gather}
\mathcal{I} = \sigma_{0}\otimes\tau_{x},\\
\mathcal{T} = i\sigma_{y}\otimes\tau_{0}\mathcal{K},
\end{gather}
where $\mathcal{K}$ is the complex conjugation operator.
 Since class D does not have the time-reversal symmetry and the spin-rotation symmetry, we break them by adding appropriate gap functions.

We consider two gap functions $\Delta_{1,2}(\mathbf{k})$,
\begin{gather}
\Delta_{1}(\mathbf{k})=\psi_{0} \sum_{n=1}^{3} e^{2n\pi i/3} \cos(\mathbf{k}\cdot\mathbf{T}_{n}) (i\sigma_{y})\otimes\tau_{0},\\
\Delta_{2}(\mathbf{k})=d_{z} \sin(2\mathbf{k}\cdot\mathbf{R}_{3})(\sigma_{x}-i\sigma_{y})(i\sigma_{y})\otimes\tau_{z}.
\end{gather}
Since the point group of the lattice is $D_{6h}$, each gap function belongs to one of the representation of $D_{6h}$.
 $\Delta_{1}(\mathbf{k})$ is a spin-singlet pairing function belonging to $E_{2g}$ representation and $\Delta_{2}(\mathbf{k})$ is a spin-triplet pairing function belonging to $E_{1u}$ representation~\cite{bzduvsek2017doubly,fischer2015symmetry}.
 
 We can easily check that $\Delta_{1}(\mathbf{k})$ breaks the time-reversal symmetry and $\Delta_{2}(\mathbf{k})$ breaks $\mathrm{U}(1)$ spin-rotation symmetry $\sigma_{z}$.
 Also, the parity of each gap function is even.
 This means that the gap function $\Delta(\mathbf{k})=\Delta_{1}(\mathbf{k})+\Delta_{2}(\mathbf{k})$ turns the tight binding Hamiltonian $\mathcal{H}(\mathbf{k})$ into the BdG Hamiltonian belonging to class D.
 And the symmetry $\mathfrak{B}$ is given by
\begin{gather}
\mathfrak{B}=s_{x}\otimes\tau_{x}\mathcal{K}.
\end{gather}
 
 We set the parameters by
\begin{gather}
E_{\mathrm{on}}=-4,~t_{1}=-1,~t_{2}=-0.5,\nonumber\\
t_{z}=-0.7,~v_{z}=0.5,\\
\psi_{0}=0.3,~d_{z}=0.5.\nonumber
\end{gather}

These parameters make tiny nodal surfaces at the edge of the first Brillouin zone which is a hexagonal prism, which are illustrated in Fig. \ref{fig:Dcharge} (a). 
 We illustrate one of the nodal surfaces at Fig. \ref{fig:Dcharge} (b).
 The red sphere is the nodal surface.
 Inside the sphere, there is a nodal points between the topmost occupied band and the second topmost occupied band, which is a blue dot in Fig. \ref{fig:Dcharge} (b).

To check the linking structure between the nodal surface and the nodal point, we first check the 0D charge of the nodal surface.
 The result of the Pfaffian calculation is illustrated at Fig. \ref{fig:Dcharge} (c).
 There is a cross section of the nodal surface at $k_{z}=0.293\cdot(2\pi/c)$. 
 The light red(white) region is where $\mathrm{Pf}[iH_{\mathrm{FB}}]$ is positive(negative). 
 Since the sign of $\mathrm{Pf}[iH_{\mathrm{FB}}]$ is changed across the nodal surface, its 0D charge is non-trivial.
 
The next step is calculating the 2D charge of the nodal surface, which is given by Eq. (\ref{2Din}). 
 We calculate the Chern number $c_{\mathrm{A},n}^{\mathrm{occ}}(S^{2})$ of the n-th occupied band using the Berry phase of the band on $S^{2}$.
 The Berry curvature $\mathbf{F}_{n}^{\mathrm{occ}}(\mathbf{k})=\nabla_{\mathbf{k}}\times \mathbf{A}_{n}^{\mathrm{occ}}(\mathbf{k})$ of the n-th occupied band is well-defined over the $S^{2}$ because the n-th occupied band is gapped from the other bands on $S^{2}$.
 Therefore, we can get $c_{\mathrm{A},n}^{\mathrm{occ}}(S^{2})$ by calculating a surface integral of $\mathbf{F}_{n}^{\mathrm{occ}}(\mathbf{k})$ over a pierced $S^{2}$, i.e. $S^{2}-\{x_{1},\cdots,x_{n}\}$ where $x_{1},\cdots,x_{n}\in S^{2}$.
 Let us pierce $S^{2}$ in its north pole and south pole.
 Using the Stokes' theorem,  we can get the below equation, 
\begin{gather}
c_{\mathrm{A},n}^{\mathrm{occ}}(S^{2})=\frac{1}{2\pi}\left(\gamma_{n}(S_{\theta =\pi}^{1})-\gamma_{n}(S_{\theta =0}^{1})\right),\label{ChernBerry}
\end{gather}
where $S^{1}_{\theta=0(\pi)}$ is a tiny circle surrounding the north(south) pole and $\gamma_{n}(S^{1})$ is the Berry phase of the n-th occupied band along $S^{1}$ which is given by
\begin{gather}
\gamma_{n}(S^{1})=i\oint_{S^{1}} d\mathbf{l}\cdot\mathbf{A}_{n}^{\mathrm{occ}}(\mathbf{k}).
\end{gather}
 Let us denote a circle, which is a parallel of $S^{2}$ with a latitude $\theta\in[0,\pi]$, as $S^{1}_{\theta}$.
 Then the Eq. (\ref{ChernBerry}) is deformed to
\begin{gather}
c_{\mathrm{A},n}^{\mathrm{occ}}(S^{2})=\frac{1}{2\pi}\int_{0}^{\pi} d\theta\ \frac{d}{d\theta}\gamma_{n}(S^{1}_{\theta}),
\end{gather}
which means that we can calculate $c_{\mathrm{A},n}^{\mathrm{occ}}(S^{2})$ by evaluating the difference between $\gamma_{n}(S^{1}_{\theta=\pi})$ and $\gamma_{n}(S^{1}_{\theta=0})$ while $\theta$ changes from $0$ to $\pi$.

We illustrate $\gamma_{n}(S^{1}_{\theta})$ for $\theta\in[0,\pi]$ in Fig. \ref{fig:Dcharge} (d). 
 To calculate the 2D charge of the nodal surface, the Chern numbers of the occupied bands outside the nodal surface is needed.
 We display the $\gamma_{n}(S^{1}_{\theta})$ calculated outside the nodal surface as the straight lines.
 The green line corresponds to the $\gamma_{1}(S^{1}_{\theta})$ and $\gamma_{2}(S^{1}_{\theta})$ and the blue line corresponds to the $\gamma_{3}(S^{1}_{\theta})$ and $\gamma_{4}(S^{1}_{\theta})$.
 We can check that the green line is increasing and winds from $\pi$ to $-\pi$ one time.
 Therefore, the 2D charge of the nodal surface is $2$.

The Chern number of the highest occupied band inside the nodal surface can be similarly calculated.
 The dashed orange line in Fig. \ref{fig:Dcharge} (d) corresponds to $\gamma_{1}(S^{1}_{\theta})$ inside the nodal surface.
 It is decreasing and winds one time from $-\pi$ to $\pi$, which means that the Chern number of the highest occupied band inside the nodal surface is $-1$. 
 Therefore, we can check that the linking structure in class D, which is Eq. (\ref{Dlinkingeq}), is satisfied in this model.
 
\section{First-principles calculations}
To simulate electronic structure of AA-stacked bilayer graphene multilayers, we have performed density functional theory (DFT) calculation. We have used projector augmented wave band method implemented in Vienna ab initio simulation package (VASP) \cite{kresse1999ultrasoft,kresse1996efficient,kresse1996efficiency} with generalized-gradient approximation \cite{perdew1996generalized}. In-plane hexagonal lattice constant is 2.46 $\mathrm{\AA}$ and Interlayer distances are 2.5 $\mathrm{\AA}$ and 3.5 $\mathrm{\AA}$ respectively, as indicated in Fig. 4 of the main text.

\end{document}